\documentclass[conference]{IEEEtran}

 \usepackage{hyperref}

\usepackage{amsmath,amssymb,graphicx,subfigure, mathtools, relsize}
\usepackage{color}

\begin{document}

\title{Geometry-constrained Degrees of Freedom Analysis for Imaging Systems: Monostatic and Multistatic}

\author{\IEEEauthorblockN{B. Mamandipoor$^{(1)}$, A. Arbabian$^{(2)}$ and U. Madhow$^{(1)}$ }\\\vspace{-0.1cm}
\IEEEauthorblockA{(1) ECE Department, University of California, Santa Barbara, CA, USA\\
(2) EE Department, Stanford University, Stanford, CA, USA\\
Emails: \{bmamandi, madhow\}@ece.ucsb.edu,
arbabian@stanford.edu}}

\newcommand{\br}{\boldsymbol{r}}
\newcommand{\by}{\boldsymbol{y}}
\newcommand{\bz}{\boldsymbol{z}}
\newcommand{\bq}{\boldsymbol{q}}
\newcommand{\bs}{\boldsymbol{s}} 
\newcommand{\bgamma}{\boldsymbol{\gamma}} 
\newcommand{\ba}{\boldsymbol{a}}
\newcommand{\bk}{\mathbf{k}}
\newcommand{\Tmono}{\mathcal{T}_{\text{mono}}}
\newcommand{\Tmulti}{\mathcal{T}_{\text{multi}}}

\newtheorem{theorem}{Theorem}
\newtheorem{lemma}{Lemma}
\newtheorem{proposition}{Proposition}
\newtheorem{remark}{Remark}
\newtheorem{example}{Example}
\newtheorem{note}{Note}
\newtheorem{definition}{Definition}

\maketitle

\begin{abstract}
In this paper, we develop a theoretical framework for analyzing the measurable information content of an unknown scene through an active electromagnetic imaging array. 
We consider monostatic and multistatic array architectures in a one-dimensional setting. Our main results include the following: (a) we introduce the space-bandwidth product (SBP), and show that, under the Born approximation, it provides an accurate prediction of the number of the degrees of freedom (DoF) as constrained by the geometry of the scene and the imaging system; (b) we show that both monostatic and multistatic architectures have the same number of DoF; (c) we show that prior DoF analysis based on the more restrictive Fresnel approximation are obtained by specializing our results; (d) we investigate matched-filter (back-propagation) and pseudoinverse image reconstruction schemes, and analyze the achievable resolution through these methods. Our analytical framework opens up new avenues to investigate signal processing techniques that aim to reconstruct the reflectivity function of the scene by solving an inverse scattering problem, and provides insights on achievable resolution. For example, we show that matched-filter reconstruction leads to a significant resolution loss for multistatic architectures.

\end{abstract}
\IEEEoverridecommandlockouts

\IEEEpeerreviewmaketitle

\section{Introduction}
\label{sec:introduction}
The evaluation of the amount of information in an unknown scene (object) that can be inferred through measurements of radiated (or back-scattered) electromagnetic fields, is a fundamental problem that has relevance across different fields including optics \cite{Di_Fracia_DoF_image, Miller_optical}, wireless communications \cite{Miller_comm, Gruber_info_theory, MIMO}, and electromagnetic imaging \cite{Babak_IMS16}. One of the crucial measures of the information capacity of such systems is the number of {\it degrees of freedom} (DoF). In general, a scene can be described by an infinite number of independent parameters. However,  the number of independent parameters that can be measured through an imaging system is typically finite \cite{Di_Francia_resolving_power}, and is given by the number of DoF of the system. DoF analysis is of both theoretical and practical significance: (i) it is related to fundamental performance measures such as achievable resolution and the information capacity of the system, (ii) it provides guidelines to design efficient and practical array architectures under various cost and complexity constraints, (iii) it provides crucial insights on the performance of different image reconstruction algorithms. 

While our focus here is on theoretical limits, we are particularly interested in imaging at high carrier frequencies (e.g., using millimeter wave bands), where synchronization across large baselines is challenging, and implementing a multistatic array presents difficulties in terms of both cost and complexity.  
In this paper, we consider two canonical active imaging array architectures, {\it monostatic} and {\it multistatic} \cite{Ahmed_magazine}. A monostatic array consists of standalone transceiver (TRx) elements, i.e., when one of the elements is transmitting, only the co-located receiver collects the back-scattered  signal. On the other hand in a multistatic architecture, for any transmitting element, all of the receivers across the array collect the back-scattered signal. In order to implement a multistatic array, we need to {\it synchronize} all of the TRx elements across the aperture. For imaging at high carrier frequencies (e.g., using millimeter wave and THz bands), which offer the potential for
significantly improved resolution, synchronization across large baselines is challenging, so that implementing a multistatic array presents significant difficulties in terms of both cost and complexity.  On the other hand, it is known that a multistatic architecture is capable of measuring a greater portion of the $k$-space spectrum for any point scatterer in the scene. It is therefore of significant interest to understand, at a fundamental level, potential improvements in information capacity (measured in terms of DoF) that a multistatic
architecture might be able to provide.


We formulate the problem under the ``Born approximation'' \cite{Tsang_book}, which is based on a {\it weak scattering} model where the total electromagnetic field at the scene is approximated by the incident field. Under this assumption, the measurement model is linearized, hence we can resort to singular value decomposition (SVD) to analyze the model. 
Additionally, we present a theory for DoF evaluation of narrowband (single frequency) 1-dimensional (1D) monostatic and multistatic imaging systems, and provide guidelines for system design and performance evaluation of image reconstruction techniques. The ideas and results presented in this paper can  be generalized to 2-dimensional (2D) and wideband (multi-frequency) imaging systems. Detailed analysis for such settings is beyond the scope of the present paper.

\subsection{Contributions}
Our key contributions are as follows:
\begin{enumerate}
\item  We introduce  the {\it space-bandwidth product} (SBP), defined by the product of the scene area and the measurable Fourier extent of the scene as ``seen'' by the imaging system (after removing redundancies), as an approximation to the DoF. SBP can be thought of as a generalization of the so-called Shannon number \cite{Shannon_number} for  spectral measurements of an unknown scene through a space-variant bandlimited system. We evaluate the accuracy of our proposed DoF measure by comparing it to numerical SVD computations for various geometries.

\item We specialize our SBP analysis to deduce prior results that had been derived for parallel planar surfaces using a more restrictive Fresnel approximation. We also provide a clear analysis of the design guidelines for multistatic arrays that are based on the ``effective aperture'' concept.

\item We investigate image formation techniques that aim to reconstruct the reflectivity function of the scene by solving an inverse scattering problem. We show that the SVD analysis provides an easy understanding of the measurement process, as well as the achievable resolution of various reconstruction schemes.  In particular, we show that back-propagation reconstruction for multitstaic architecture is highly suboptimal and leads to a significant loss in image resolution. 

\end{enumerate}

\subsection{Related Work}
Classical theories for DoF analysis of imaging systems are based on the Shannon sampling theorem. In a series of fundamental papers, Di Francia derived the significant conclusion that an image formed through a finite size aperture has a finite number of DoF \cite{Di_Francia_resolving_power, Di_Francia_supergain}. Since there is no limitation on the number of DoF of the scene, it follows that many different scenes can map to exactly the same image.  However, this result is not mathematically correct, as has been pointed out by multiple authors \cite{Wolter_book, Slepian_bandwidth}. The reason is that, if the scene is of finite size, then the knowledge of its Fourier transform over a bounded domain is enough to reconstruct it exactly by using analytic continuation. In order to account for the inevitable existence of noise in practical systems, it is necessary to introduce the notion of {\it effective,} or {\it practically useful,} DoF \cite{Di_Fracia_DoF_image}. This can be accomplished by applying the seminal work of Slepian, Landau, and Pollak on prolate spheroidal wave functions (PSWFs) \cite{Slepian, PSWF_three}. The PSWF theory shows that the eigenvalues corresponding to a finite Fourier integral operator are approximately constant up to a critical point, after which they decay exponentially to zero. Consequently, in the presence of noise, only a finite number of eigenvalues can be  used to accurately determine the output of the integral operator.

The PSWF theory can be directly applied to the geometry of symmetric parallel planar surfaces in the far field, where we can use  the Fresnel approximation \cite{Lee_book, Frieden_PSWFs} for analyzing the measurement model \cite{Pierri_heuristic, Pierri_Resolution_2D, Solimene_Strip_Sources}. It has been shown using this approach that the number of DoF of such imaging systems is also finite, and that the corresponding eigenfunctions are related to the PSWFs. Similar techniques have been applied to analyze multiple observation domains \cite{Pierri_2_domains} and orthogonal planes geometry \cite{Pierri_depth}. However, these results do not directly generalize to short range (where the Fresnel approximation is not accurate), or to asymmetric or tilted planes geometries. The authors in \cite{Miller_optical} present a general theory for {\it computing} the electromagnetic DoF of optical systems under arbitrary boundary conditions, following a sequential optimization framework for finding the strongest connected source and receiving functions that span the input and output spaces, respectively. This approach is equivalent to the SVD of the system integral operator, and it shows that the number of practically useful DoF are essentially finite under general boundary conditions. In this paper, we  propose SBP as a measure to predict the number of DoF, and use numerical SVD computations to verify the accuracy of our measure for various geometries. 

The term space-bandwidth product has been previously used in a somewhat different context, for evaluating the information content of optical signals and systems \cite{Lohmann_SBP, Lohmann_SBP_2}.  In this setting, it is defined as the area within the Wigner distribution representation of a signal. The latter forms an 
intermediate signal description between the pure spatial representation and the pure Fourier domain representation, and can be roughly interpreted as the local spatial spectrum of the signal \cite{Bastiaans_Wigner}.  The Wigner distribution has been derived for a 1D signal and its corresponding 1D Fourier Transform, and does not apply to our measurement model, where the desired information is seen through an active electromagnetic imaging system. In our definition of space-bandwidth product, the (spatial) ``bandwidth'' is the $k$-space spectrum for each point scatterer in the scene, which depends on the imaging system. This is then integrated over all of the point scatterers constituting the scene, or ``space,'' to obtain the space-bandwidth product.  Thus, space-bandwidth product as defined in this paper depends on the geometry of both the imaging system and the scene.


\subsection{Organization}
Section \ref{sec:background} presents the measurement model and mathematical background for SVD analysis and the $k$-space representation of the system. Sections \ref{sec:k_space_spectrum} investigates the $k$-space spectrum corresponding to a point scatterer seen through monostatic and multistatic array architectures. In Section \ref{sec:SBP}, we go through the details of SBP computations for different imaging geometries, and verify the accuracy of the results by numerical SVD computations. Section \ref{sec:Fresnel} investigates the implications of the the Fresnel approximation for parallel planes geometry; it shows that SBP computations converge to approximate solutions of previous models, and provides insights for the design of multistatic arrays in the Fresnel regime. In Section \ref{sec:reconstruction} we investigate image formation techniques that aim to solve the inverse scattering problem, and analyze the achievable resolution for monostatic and multistatic arrays. Finally, our conclusions are presented in Section \ref{sec:conclusions}.

\section{Mathematical Model and Background}
\label{sec:background}

\begin{figure}[htbp]
\centering
\includegraphics[scale=0.65] {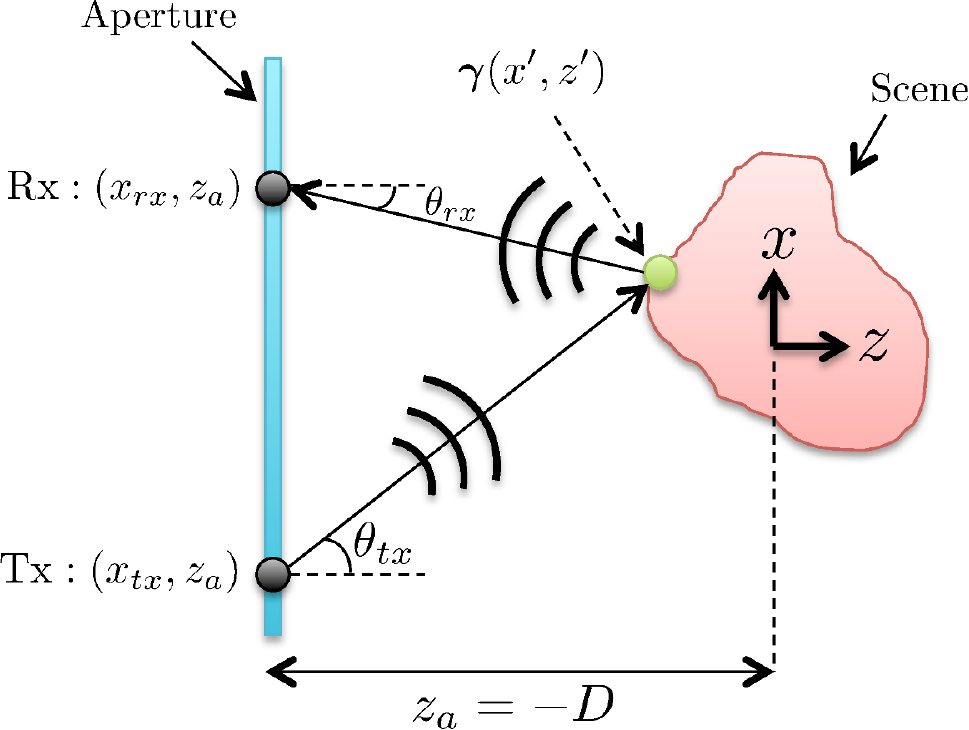}
  \caption{Geometry of 1-dimensional bistatic pair at distance $z_{a}=-D$ from the center of the scene.}
  \label{fig:math_model} 
\end{figure}

Consider the 1D aperture  depicted in Figure \ref{fig:math_model}. The scene (object) is located at the origin, with the corresponding reflectivity function defined by $\bgamma: \mathcal{A}\to\mathbb{C}$, where $\mathcal{A}\subset \mathbb{R}^{2}$ is bounded.
We restrict the Tx and Rx antenna elements to be located on the same plane: $z_a = -D$.
Assume that the Tx element located at $(x_{tx},z_a)$ illuminates the entire scene, and that the Rx element located at $(x_{rx} , z_{a})$, measures the back-scattered signal. The observed signal over an aperture of length $L_1$ is denoted by $\bs: \mathcal{B}\to \mathbb{C}$, where $\mathcal{B} = [-L_{1}/2, L_{1}/2]\times[-L_{1}/2, L_{1}/2]$. The relationship between the scene reflectivity function and the observed signal over the aperture is governed by the Helmholtz wave equation \cite{Cheng_book}. Under the Born approximation \cite{Tsang_book}, the solution of the scalar Helmholtz  equation for homogenous isotropic media (simplified by dropping space attenuation factors) satisfies the following {\it linear} integral equation:
\begin{equation}
\label{eq:born}
\bs(x_{tx}, x_{rx}) =\!\! \int\limits_{\mathcal{A}} \xi(x_{tx}, x_{rx}, x',z')\bgamma(x',z')  dx' dz',
\end{equation}
where
\begin{equation}
\label{eq:zeta}
\xi (x_{tx}, x_{rx}, x',z') = e^{-jkR(x_{tx}, x';z')} e^{-jkR(x_{rx},x';z')},
\end{equation}
denotes the space-variant impulse response of the system, and $R(x,x';z') \triangleq \sqrt{(x-x')^2 + (z_a - z')^2}$ is the Euclidean distance between the point $(x,z_a)$ on the aperture plane, and the point scatterer in the scene located at $(x',z')$. The wavenumber is denoted by $k = {2\pi\over \lambda}$, where $\lambda$ is the signal wavelength. We assume $\bgamma\in\Psi$ and $s\in\Phi$, where  $\Psi \triangleq \mathcal{L}^{2}(\mathcal{A})$ and $\Phi \triangleq \mathcal{L}^{2}(\mathcal{B})$ represent the Hilbert spaces of square integrable functions over $\mathcal{A}$ and $\mathcal{B}$, respectively. 
This assumption places the (physically plausible) restriction that the scene reflectivity function and the scattered electromagnetic fields have finite energy values.  It is convenient to recast the linear observation model in the following  operator form:
\begin{equation}
\label{eq:Xi}
\bs = \Xi \bgamma,
\end{equation}
where $\Xi:\Psi\to\Phi$  is defined by the right-hand side of  (\ref{eq:born}). It is easy to see that the integral kernel (\ref{eq:zeta}) satisfies 
\begin{equation}
\label{eq:zeta_square_integrable}
||\Xi||_{HS}^2\triangleq \iint\limits_{\mathcal{A}\mathcal{B}} |\xi(x_{tx}, x_{rx}, x',z')|^{2} dx_{tx}dx_{rx}dx' dz' < \infty,
\end{equation}
where $||\Xi||_{HS}$ denotes the Hilbert-Schmidt norm (extension of Frobenius norm to possibly infinite dimensional Hilbert spaces) of the integral operator $\Xi$. Therefore, $\Xi$ belongs to the class of Hilbert-Schmidt operators, and is {\it compact} \cite{Bertero_book}. By virtue of linearity and compactness, we can invoke the Spectral Theorem \cite{Bertero_book, Porter_book}, and introduce the singular value decomposition of $\Xi$, denoted by $\{\sigma_{i}, \psi_{i}, \phi_{i}\}_{i\in\mathbb{N}}$, where $\sigma_{i}\in\mathbb{R}^{+}$are the singular values, and $\psi_{i}\in \Psi$ and $\phi_{i}\in \Phi$ are the right and left singular functions, respectively. The operator $\bs = \Xi\bgamma $ can therefore be expressed as
\begin{equation}
\label{eq:Xi_decomposition}
\bs = \sum_{i=1}^{\infty} \sigma_{i} \phi_{i} \langle \bgamma , \psi_{i}\rangle_{\Psi} ,
\end{equation}
where $\langle \cdot , \cdot \rangle_{\Psi} $ denotes the inner product in $\Psi$. The SVD of $\Xi$ is equivalent to the following series expansion of the integral kernel in (\ref{eq:zeta}): 
\begin{equation}
\label{eq:zeta_decomposition} 
\xi (x_{tx},x_{rx}, x',z') = \sum_{i = 1}^{\infty} \sigma_{i} \phi_{i}(x_{tx},x_{rx}) \psi^{*}_{i}(x',z').
\end{equation}
The sets of singular functions $\{\psi_{i}\}_{i\in\mathbb{N}}$ and $\{\phi_{i}\}_{i\in\mathbb{N}}$ are orthonormal bases for $\Psi$ and $\Phi$, respectively. From (\ref{eq:Xi_decomposition}), we obtain the following one-to-one correspondence between the two sets of singular functions,
\begin{equation}
\phi_{i} = {1\over \sigma_i} \Xi \psi_{i}, \quad \forall i.
\end{equation}

\subsection{Sum rule for computing the operator norm}
Using Parseval's Theorem, it is easy to see the following sum rule \cite{Miller_optical, Miller_comm} (see Appendix \ref{app:sum_rule} for the proof),
\begin{equation}
\label{eq:sum_rule}
||\Xi||_{HS}^2 = \sum_{i =1}^{\infty} \sigma_i^2,
\end{equation}
i.e., the square of the operator norm is equal to the {\it sum of strengths} of its SVD components. Square integrability of the kernel in Equation (\ref{eq:zeta_square_integrable}) leads to
\begin{equation}
 \sum_{i = 1}^{\infty} \sigma_{i}^{2} < \infty,
\end{equation}
that is, the sum of squares of the singular values of a Hilbert-Schmidt operator converges \cite{Bertero_book}.
Therefore, putting the sequence of singular values in non-increasing order, we have $\sigma_{i}^{2}\to 0$ for $i\to\infty$.
In other words, although in principle the number of nonzero singular values could be infinite, the number of {\it practically useful} singular values is finite \cite{Di_Fracia_DoF_image, Miller_optical}. The normalized sum of singular values $\bar{\Sigma} = \sum_{i}(\sigma_{i}/\sigma_{max})$ \cite{Shannon_number, PSWF_three}, or the normalized sum of squares of the singular values $\bar{\Sigma}_{sq} = ||\Xi||_{HS}^2/\sigma_{max}^2$ \cite{Miller_optical}, where $\sigma_{max} = \sigma_1 = \max\{\sigma_i\}$, can be associated with the number of degrees of freedom for a system with a {\it steplike} behavior of the singular values , where all useful nonzero singular values are approximately equal up to a certain threshold, after which they rapidly decay to zero. Note that for computing $\bar{\Sigma}_{sq}$ we can sidestep the SVD computation by using (\ref{eq:zeta_square_integrable}) and only computing the maximum singular value of the operator. In general, this steplike behavior is not satisfied for imaging systems, and the singular values decay gradually to zero \cite{Miller_optical}. A key goal in this paper is to find a simple criterion to determine the number of useful nonzero singular values of $\Xi$ for different imaging scenarios.

\noindent {\bf Invariance of the operator norm to the relative geometry:} The integral kernel in (\ref{eq:zeta}) has unit magnitude over its entire parameter space. Therefore, the operator norm in (\ref{eq:zeta_square_integrable}) reduces to,
\begin{equation}
||\Xi||_{HS}^2 = V_{\mathcal{A}} V_{\mathcal{B}},
\end{equation}
where $V_{\mathcal{A}}$, $V_{\mathcal{B}}$ represent the volume of the sets $\mathcal{A}$ and $\mathcal{B}$, respectively. This observation indicates that for the measurement model in (\ref{eq:born}), the operator norm as well as the dependent criteria for estimating the degrees of freedom (e.g., $\bar{\Sigma}_{sq}$), are invariant to the relative geometry of the aperture and the scene, and hence are not capable of capturing the effect of rotation and translation of the scene (or the aperture) on the available degrees of freedom of the imaging system. 
In general, the number of independent parameters that we can extract from an unknown object using a linear operator is precisely determined by the number of nonzero singular values of the operator. For our simulations we compute the singular system of the operator $\Xi$ by discretizing the kernel provided by equation (\ref{eq:zeta}), over the parameter spaces $\mathcal{A}$ and $\mathcal{B}$, and compute the SVD numerically. 
In the next subsection, we review the $k$-space (spatial frequency domain) representation of the integral operation in Equation (\ref{eq:born}), which is a crucial step in defining the SBP of the imaging system.

\subsection{$k$-space representation}
Taking the $2$D Fourier Transform (FT) of $\bs(x_{tx}, x_{rx})$ of the received signal given by (\ref{eq:born}) and (\ref{eq:zeta}) over the aperture, yields the data representation in the spatial frequency domain,
\begin{eqnarray}
&&S(k_{x_{tx}} , k_{x_{rx}}) = \text{FT}_{2D}\{\bs(x_{tx}, x_{rx})\} \notag\\
&\triangleq& \iint\limits_{x_{tx}  x_{rx}} \bs(x_{tx} , x_{rx}) e^{-jk_{x_{tx}}x_{tx}} e^{-jk_{x_{rx}}x_{rx}} dx_{tx} dx_{rx}.
\end{eqnarray}
Substituting the expression for $\bs(x_{tx}, x_{rx})$ from (\ref{eq:born}), and changing the order of integration yields 
\begin{eqnarray}
\label{eq:2d_ft}
S(k_{x_{tx}} , k_{x_{rx}}) &=& \notag \\
 \iint\limits_{z'x'} \bgamma(x',z') && \!\!\!\!\!\!\!\!\!\!\!\!\!\!\! \tilde{\xi}(k_{x_{tx}}, k_{x_{rx}}, x',z') dx' dz',
\end{eqnarray}
where $\tilde{\xi}(k_{x_{tx}}, k_{x_{rx}}, x',z') = \text{FT}_{2D}\{ \xi(x_{tx}, x_{rx}, x', z')\}$ denotes the space-variant transfer function of the system. The $2$D FT operator can be decomposed into two $1$D Fourier Transforms as follows:
\begin{eqnarray}
\label{eq:xi_freq_domain}
&&\!\!\!\!\!\!\!\!\!\!\!\!  \tilde{\xi}(k_{x_{tx}}, k_{x_{rx}}, x',z') = \notag\\
&&\text{FT}_{1D}\{ e^{-jkR(x_{tx}, x';z')}\}  \text{FT}_{1D}\{ e^{-jkR(x_{rx}, x';z')}\},
\end{eqnarray}
where $\text{FT}_{1D}\{f(\alpha)\} \triangleq \int_{\alpha} f(\alpha) e^{-jk_{\alpha} \alpha} d\alpha$. 

We now estimate the $1$D FTs in (\ref{eq:xi_freq_domain})
using the method of stationary phase \cite{Chew_book, Ahmed_thesis}, which provides an  approximate solution for integrals of oscillatory functions. The method involves determining the points where the phase is stationary (i.e., where the derivative equals zero), and replacing the integral with the sum of the function values at the stationary points. 
Thus, an approximate solution for the 1D FTs in (\ref{eq:xi_freq_domain}) is given by 
\begin{eqnarray}
\label{eq:FT_by_stationary_phase}
\!\!\!\! \text{FT}_{1D}\{ e^{-jkR(x_{tx}, x';z')}\} &\approx& \!\!\!\! e^{-jkR(x_{tx}^{sp},x';z')}e^{-jk_{x_{tx}}x_{tx}^{sp}}\notag\\
\!\!\!\! \text{FT}_{1D}\{ e^{-jkR(x_{rx}, x';z')}\} &\approx& \!\!\!\! e^{-jkR(x_{rx}^{sp},x';z')}e^{-jk_{x_{rx}}x_{rx}^{sp}}.
\end{eqnarray}
where $x_{tx}^{sp}$ and $x_{rx}^{sp}$ are stationary points in the Fourier integral arguments, which are easily shown to satisfy
\begin{eqnarray}
\label{eq:stationary_phase}
k_{x_{tx}} R(x_{tx}^{sp}, x';z')  &=& k (x' - x_{tx}^{sp}) \notag\\
k_{x_{rx}} R(x_{rx}^{sp}, x';z')  &=& k (x' - x_{rx}^{sp}).
\end{eqnarray}
We now utilize these expressions to reduce (\ref{eq:FT_by_stationary_phase}) into a function of $x'$ and $z'$. In the process, we provide a geometric interpretation of the stationary phase approximation.

\begin{figure}[htbp]
\centering 
\subfigure[]{
\includegraphics[scale=0.60]{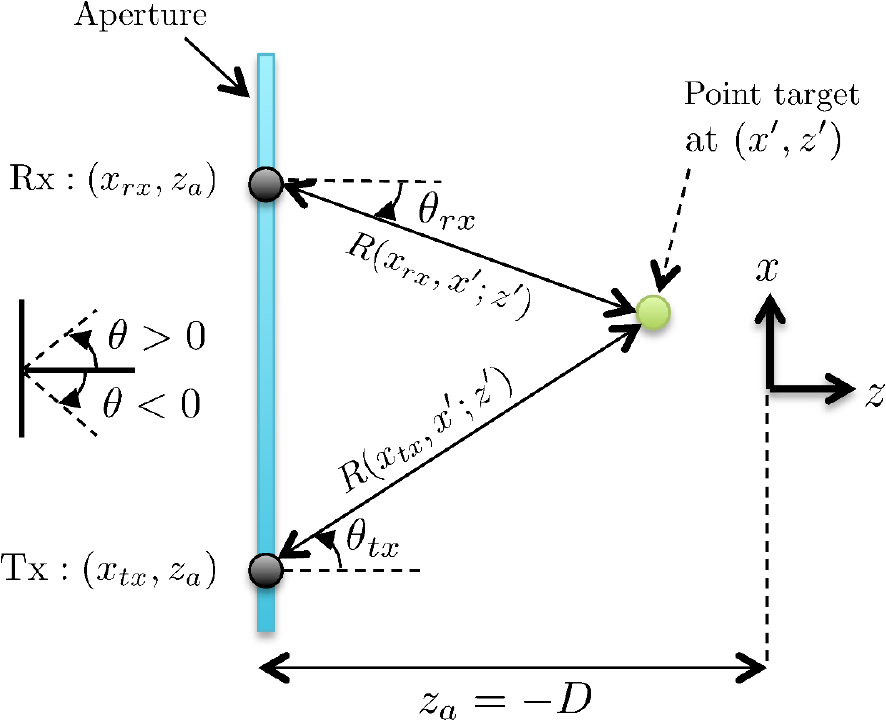}
}
\subfigure[]{
\includegraphics[scale=0.39]{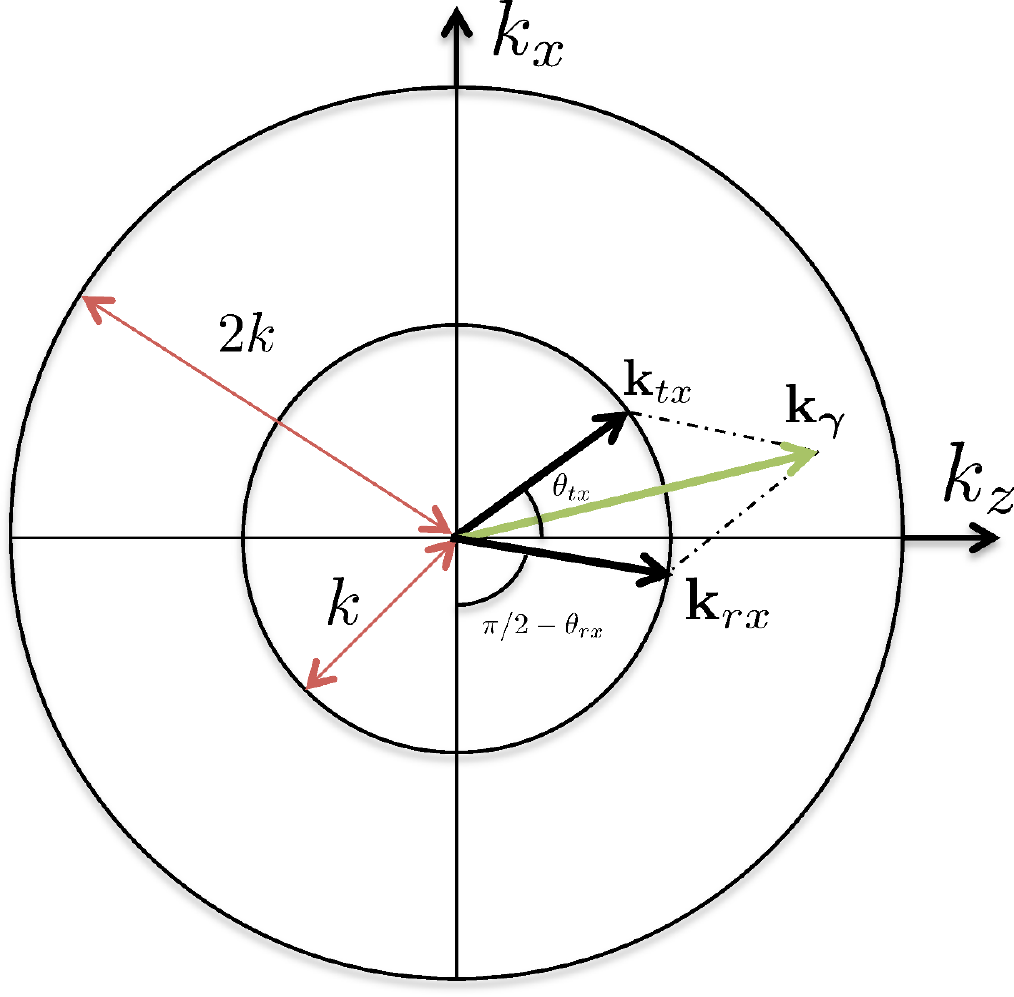}
}
\caption{(a) The geometry of bistatic Tx/Rx elements and a point scatterer in the scene, (b) the corresponding sampled point in the spectrum of the point scatterer in $k$-space.}
\label{fig:math_model_k_space} 
\end{figure}

\noindent \textbf{Geometric interpretation of stationary phase approximation:} Consider a given point scatterer located at $(x',z')$, and a given bistatic Tx/Rx pair, depicted in Figure \ref{fig:math_model_k_space}-a. The stationary phase condition (\ref{eq:stationary_phase}) simply characterizes the spatial frequency components corresponding to the dominant propagating plane waves for this geometry.  In particular, it can be rewritten as\begin{equation} \label{eq:kx}
k_{x_{tx}} = k \sin(\theta_{tx}), \quad k_{x_{rx}} = k \sin(\theta_{rx}).
\end{equation}
Substituting this into the phase term for the first equation in (\ref{eq:FT_by_stationary_phase}), we obtain
\begin{eqnarray} \label{eq:stationary_simplify}
&&kR(x_{tx}^{sp},x';z') + k_{x_{tx}}x_{tx}^{sp}  \notag\\
&=&k(z' - z_a)/ \cos(\theta_{tx}) + k_{x_{tx}}\left( x' - (z'-z_a)\tan(\theta_{tx})\right) \notag\\
&=&k_{x_{tx}}x' + k\cos(\theta_{tx})(z'-z_a)  \notag\\
&=&k_{x_{tx}}x' + k_{z_{tx}}(z'-z_a),
\end{eqnarray}
where it becomes natural to define $k_{z_{tx}} = k \cos(\theta_{tx})$ as the dominant spatial frequency in the $z$ direction for the Tx.   An entirely similar computation and interpretation holds for the second equation in (\ref{eq:FT_by_stationary_phase}).  

To summarize, the stationary phase approximation corresponds to propagation along the dominant spatial frequencies along the $x$ and $z$ directions for the Tx and Rx, given in terms of viewing angles shown in Figure \ref{fig:math_model_k_space}-a as follows: 
\begin{eqnarray}
\label{eq:viewing_angle}
k_{x_{tx}} = k \sin(\theta_{tx}), \quad k_{z_{tx}} = k \cos(\theta_{tx}) =  \sqrt{k^2 - k_{x_{tx}}^2}, \notag\\
k_{x_{rx}} = k \sin(\theta_{rx}), \quad k_{z_{rx}} = k \cos(\theta_{rx}) =  \sqrt{k^2 - k_{x_{rx}}^2}. 
\end{eqnarray} 

We can now write the 1D Fourier transforms in (\ref{eq:FT_by_stationary_phase}) as
\begin{eqnarray}
\label{eq:1d_ft}
\text{FT}_{1D}\{ e^{-jkR(x_{tx}, x';z')}\} &\approx& e^{-jk_{z_{tx}}(z'-z_a)} e^{-jk_{x_{tx}}x'} \notag\\
\text{FT}_{1D}\{ e^{-jkR(x_{rx}, x';z')}\} &\approx& e^{-jk_{z_{rx}}(z'-z_a)} e^{-jk_{x_{rx}}x'}, 
\end{eqnarray}
Substituting (\ref{eq:xi_freq_domain}) and (\ref{eq:1d_ft}) in (\ref{eq:2d_ft}) yields the following approximation for the 2D spectrum:
\begin{eqnarray}
\label{eq:sample_gamma}
&&S(k_{x_{tx}} , k_{x_{rx}}) \approx  e^{j(k_{z_{tx}}+k_{z_{rx}})z_{a}}  \times \notag\\
&&\iint\limits_{z'x'} \bgamma(x',z') e^{-j(k_{x_{tx}} + k_{x_{rx}})x'} e^{-j(k_{z_{tx}} + k_{z_{rx}})z'}   dx'dz'  \notag\\
&& = e^{j(k_{z_{tx}}+k_{z_{rx}})z_{a}}  \tilde{\bgamma}( k_{x_{tx}}+k_{x_{rx}} , k_{z_{tx}}+k_{z_{rx}} ),
\end{eqnarray}
where $\tilde{\bgamma}(k_x, k_z) \triangleq \text{FT}_{2D}\{\bgamma(x',z')\}$, is the $2$D spectrum of the scene reflectivity function.
Let us define the $2$D $k$-space vectors corresponding to the Tx and Rx as $\bk_{tx} = (k_{x_{tx}}, k_{z_{tx}})$, and $\bk_{rx} = (k_{x_{rx}}, k_{z_{rx}})$, respectively.
From equation (\ref{eq:sample_gamma}), we see that the spectrum of the scene has been sampled at
\begin{eqnarray}
\label{eq:k_gamma}
\bk_{\bgamma} \triangleq (k_x , k_z) &=& \bk_{tx}  + \bk_{rx}  \notag\\
&=& ( k_{x_{tx}}+k_{x_{rx}} , k_{z_{tx}}+k_{z_{rx}} ),
\end{eqnarray} 
That is, $\bk_{\bgamma}$ is the summation of two vectors $\bk_{tx}$ and $\bk_{rx}$, each of which lie on the ring of radius $k$.
Therefore, using (\ref{eq:k_gamma}) and (\ref{eq:viewing_angle}), we can characterize the $k$-space spectrum of a point scatterer for any given array architecture. For instance, Figure \ref{fig:math_model_k_space}-b shows the sampled point ($\bk_{\bgamma}$) in the spectrum of the point scatterer in Figure \ref{fig:math_model_k_space}-a. See \cite{Ahmed_thesis, Lee_book} for more details on the $k$-space representation of active imaging systems.

\section{$k$-space spectrum for monostatic and multistatic arrays}
\label{sec:k_space_spectrum}

\begin{figure}[htbp]
\centering
\includegraphics[scale=0.4] {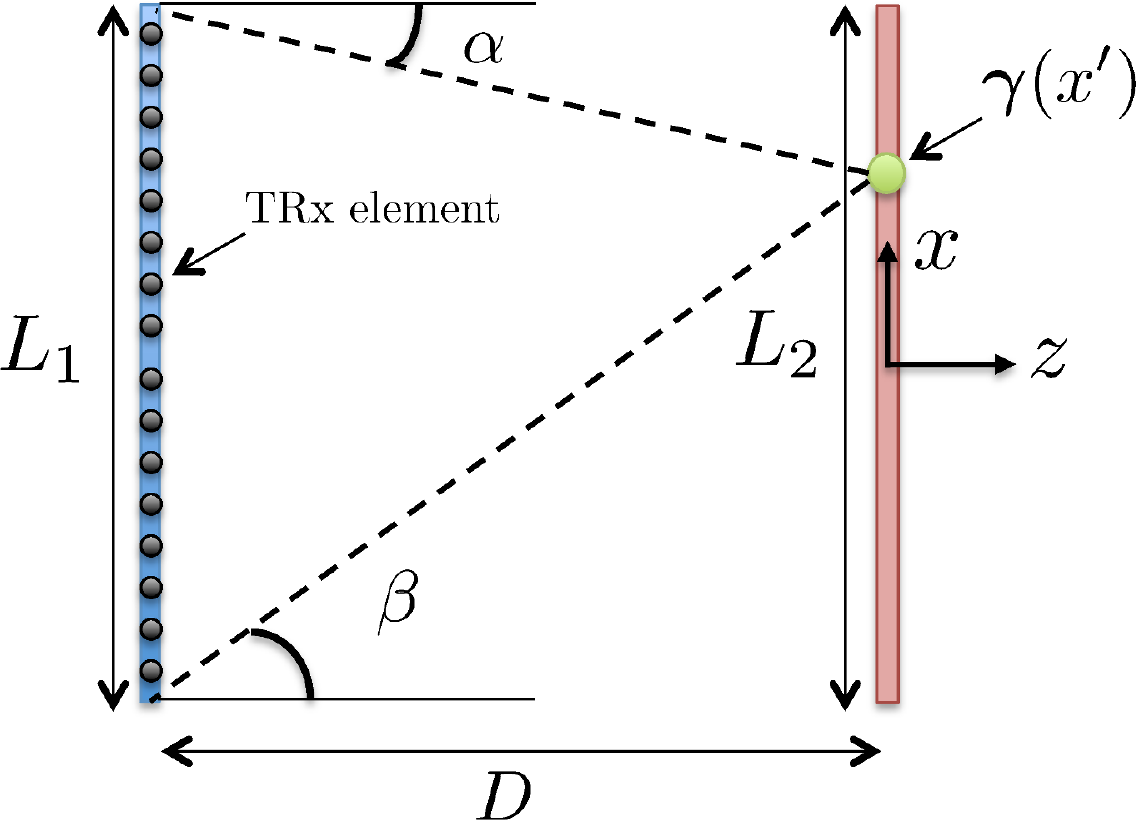}
\caption{Geometry G1: $1$D parallel and symmetric propagation model.}
\label{fig:geometry_G1} 
\end{figure}

\begin{figure}[htbp]
\centering
\includegraphics[scale=0.4] {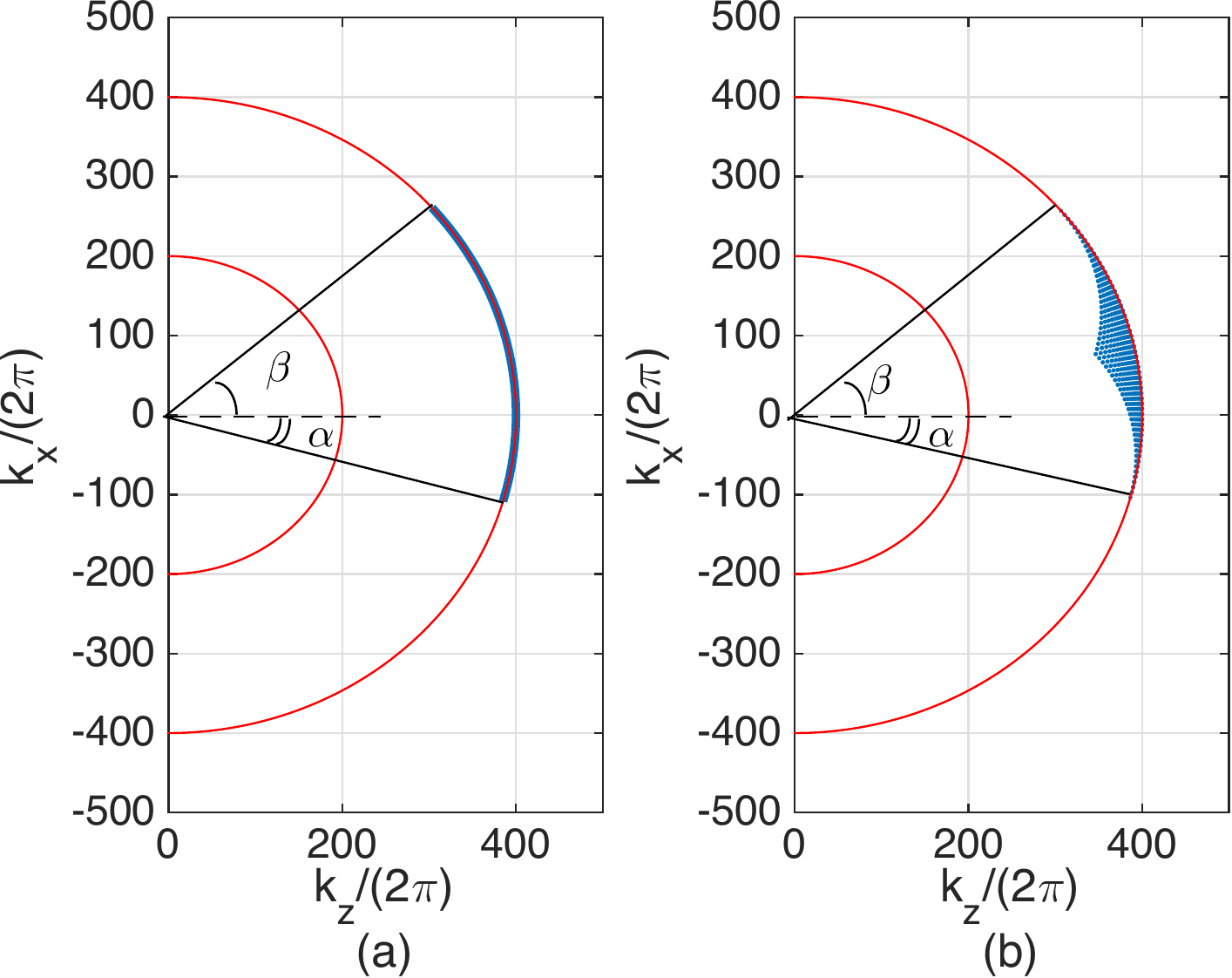}
\caption{$k$-space spectrum corresponding to point scatterer in Figure \ref{fig:geometry_G1}, seen through (a) monostatic and (b) multistatic array of infinitely many TRx elements.}
\label{fig:k_space_mono_multi} 
\end{figure}

Consider the 1-dimensional array geometry depicted in Figure \ref{fig:geometry_G1}.  Let us restrict $\bgamma(x',z')$ to a plane parallel to the aperture with reflectivity $\bgamma(x') \triangleq \bgamma(x',z'=0)$.
Let $L_1$, $L_2$, and $D$ denote the size of the aperture, the size of the scene, and the distance between the aperture and the scene, respectively. For any point scatterer located at $x=x'$ we can identify the spectral region that will be sampled using a specific array geometry. A monostatic architecture restricts the Tx and Rx to be co-located, hence, for any array element we have $\bk_{tx} = \bk_{rx}$. By equation (\ref{eq:k_gamma}), we have $\bk_{\bgamma} = 2\bk_{tx}$, i.e., the $\ell_{2}$-norm of $\bk_{\bgamma}$ is $||\bk_{\bgamma}||_{2} = 2k$, and its direction is determined by the corresponding viewing angle $\theta_{tx} = \theta_{rx}$. Figure \ref{fig:k_space_mono_multi}-a shows the spectral content (corresponding to the point scatterer in Figure \ref{fig:geometry_G1}) seen through a monostatic array of infinitely many (co-located) Tx and Rx elements. We see that $\bk_{\bgamma}$ lies on the arc of the circle of radius $2k$, confined by the angles $\alpha$ and $\beta$, the two extremes of viewing angles from the aperture (also depicted in Figure \ref{fig:geometry_G1}). Consequently, the monostatic spectrum corresponding to a point scatterer is given by
\begin{equation}
\Tmono = \{2\bk_{tx}: ||\bk_{tx}||_{2} = k,  \alpha \leq \angle \bk_{tx} \leq \beta\}.
\end{equation}

On the other hand, in a multistatic array, the Tx and Rx elements are not forced to be co-located, hence they can have different viewing angles. Figure \ref{fig:k_space_mono_multi}-b shows the spectral content of the  point scatterer seen through a multistatic array. For $\theta_{tx}\neq \theta_{rx}$,  we have
\begin{equation}
||\bk_{\bgamma}||_{2} = 2k \cos(|\theta_{tx} - \theta_{rx}|/2) < 2k.
\end{equation}
Hence, the multistatic array not only samples the region of the spectrum seen by a monostatic array, but is also able to sample points that are inside the circle of radius $2k$. Note that the {\it angular extent} of the sampled region, determined by the extreme viewing angles $\alpha$ and $\beta$, is the same for monostatic and multistatic arrays. The multistatic spectrum corresponding to a point scatterer is given by
\begin{eqnarray}
\Tmulti  &=& \{\bk_{tx} + \bk_{rx}: ||\bk_{tx}||_{2} = ||\bk_{rx}||_{2} = k, \notag\\ 
&& \alpha \leq \angle \bk_{tx} \leq \beta,  \alpha \leq \angle \bk_{rx} \leq \beta \}.
\end{eqnarray}
It is easy to see that $\Tmulti$ can also be written as,
\begin{equation}
\label{eq:T_multi}
\Tmulti = \{(p_1 + p_2)/2 : p_1, p_2 \in \Tmono\}.
\end{equation}
Therefore, 
\begin{equation}
\label{eq:T_subset_T}
\Tmono \subseteq \Tmulti \subseteq \text{conv}(\Tmono),
\end{equation} 
where $\text{conv}(\mathcal{T})$ denotes the convex hull of the set $\mathcal{T}$.

\noindent{\bf Effective monostatic: } As shown in Figure \ref{fig:math_model_k_space}, a spatially-separated pair of Tx/Rx elements samples the spectrum of a point scatterer at $\bk_{\bgamma}$, where $||\bk_{\bgamma}|| = 2k \cos(|\theta_{tx} - \theta_{rx}|/2)$, and $\angle \bk_{\bgamma} = (\theta_{tx} + \theta_{rx})/2$. Therefore, for a given point scatterer in the scene, the same information can be captured by replacing the Tx/Rx pair with a monostatic element located at $x_{eff}\in[x_{tx}, x_{rx}]$, such that $\theta_{eff} = (\theta_{tx} + \theta_{rx})/2$, and transmitting a sinusoidal wave of wavelength $\lambda_{eff} = \lambda /\cos(|\theta_{tx} - \theta_{rx}|/2)$. Note that $x_{eff}$, $\lambda_{eff}$, and the resulting effective monostatic array depends on the viewing angles for a specific point scatterer, and cannot be generalized to the entire scene. In Section \ref{sec:Fresnel}, where we investigate the Fresnel approximation (in the far field), we show that the dependence on scatterer location disappears in this regime, so that we can define an effective monostatic array that applies to the entire scene.

\begin{remark}
The computation of the operator norm for the 1-dimensional array geometry leads to $\sum_{i=1}^{\infty} \sigma_i^2$ being equal to $L_1L_2$ and $L_1^2 L_2$ for monostatic and multistatic arrays, respectively. This result holds true independent of the relative geometry of the aperture and the scene, i.e., $\sum_{i=1}^{\infty} \sigma_i^2$ remains constant with any rotation or translation of the scene or the aperture. As we see in Section \ref{sec:SBP}, the number of degrees of freedom of an imaging system depends heavily on the relative geometry of the scene and the aperture, and our proposed SBP measure is capable of capturing these dependencies for all rotation and translation parameters.
\end{remark}


\section{Space-Bandwidth Product and Degrees of Freedom}
\label{sec:SBP}
We are interested in identifying the number of DoF, or the number of independent parameters that can be extracted from an arbitrary scene, assuming that we are only constrained by the geometry of the imaging scenario. In this section, we introduce  {\it space-bandwidth product} (SBP), defined by the product of the scene area and the measured spectral extent of the scene (after removing redundancies), as a means of identifying the DoF of the system. SBP can be thought of as a generalization of the so-called Shannon number \cite{Shannon_number}, for  spectral measurements of an unknown scene through a space-variant bandlimitted system. We evaluate the accuracy of SBP measure by comparing it to numerical SVD computations for various geometries.

\begin{figure}[htbp]
\centering
\includegraphics[scale=0.4] {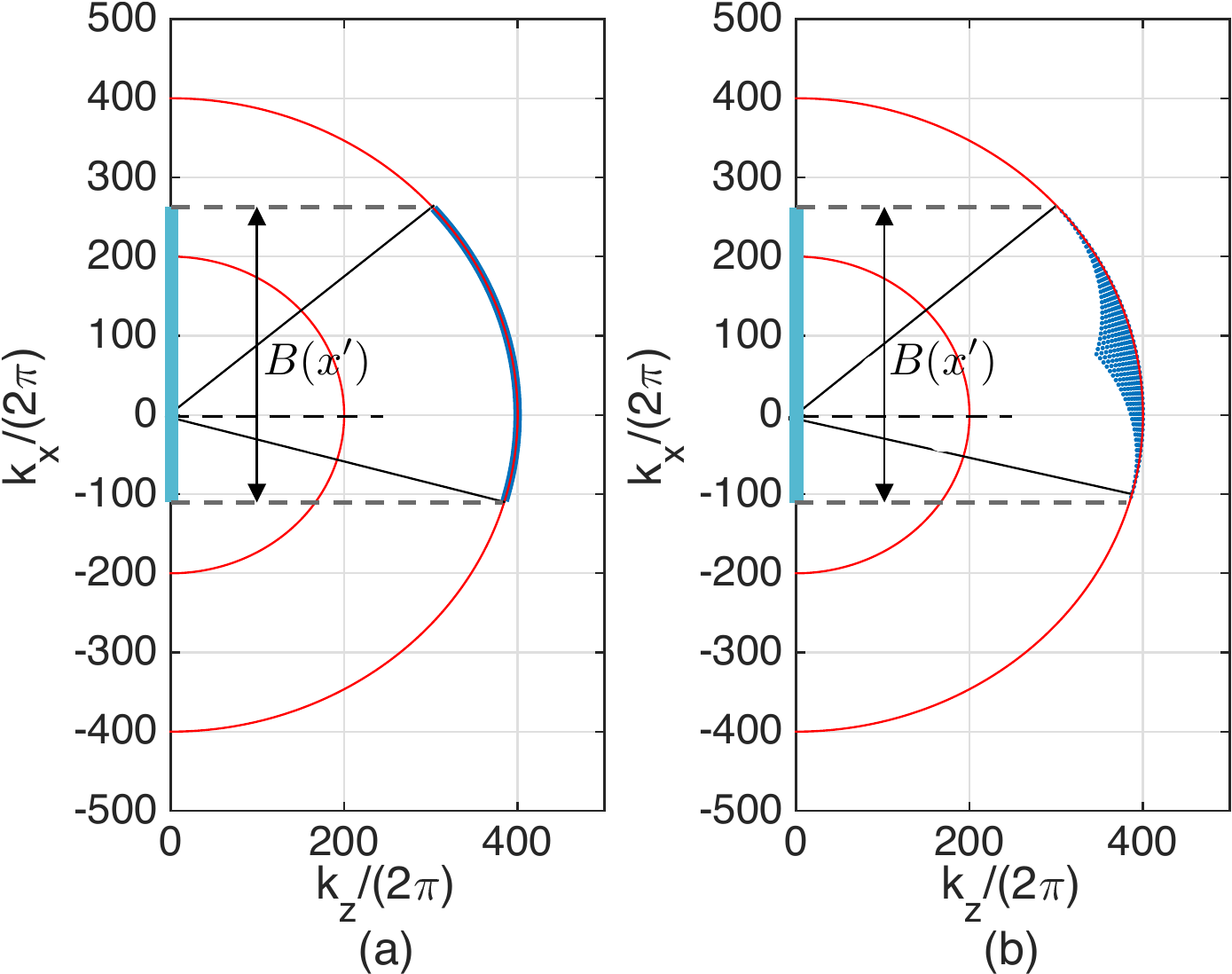}
  \caption{Spatial frequency bandwidth $B(x')$ corresponding to point scatterer in Figure \ref{fig:geometry_G1}, computed after projection of the spectrum onto $k_x$ axis for (a) monostatic and (b) multistatic array of infinitely many TRx elements.}
  \label{fig:bandwidth} 
\end{figure}

The scene information lies in $\bgamma(x')$.  We do not constrain the reflectivity function, allowing it to take any complex value for each position $x'$. Substituting $\bgamma(x') = \bgamma(x', z' = 0)$ in (\ref{eq:sample_gamma}) gives  
\begin{eqnarray}
\label{eq:sample_1d}
S(k_{x_{tx}} , k_{x_{rx}}) \!\!\!\! &=&  \!\!\!\!e^{j(k_{z_{tx}}+k_{z_{rx}})z_{a}}  \int\limits_{x'} \bgamma(x') e^{-j(k_{x_{tx}} + k_{x_{rx}})x'}  dx'\notag\\
&=& e^{jk_{z}z_{a}}  \int\limits_{x'} \bgamma(x') e^{-jk_{x}x'}  dx'.
\end{eqnarray}
Note that the integral kernel in (\ref{eq:sample_1d}) only depends on $k_{x} = k_{x_{tx}} + k_{x_{rx}}$. That is, any pair of Tx/Rx elements that leads to the same spatial frequency in the $x$ direction captures the exact same information about the scene. Hence, in order to avoid redundancy in the acquired information, we consider the {\it projection} of the sampled points in the spectrum onto the $k_{x}$ axis. For any point scatterer located at $x = x'$, let us define the {\it spatial frequency bandwidth} $B(x')$ as the width of the corresponding spectrum after the projection operation. Figure \ref{fig:bandwidth} shows $B(x')$ corresponding to the point scatterer in Figure \ref{fig:geometry_G1}, for monostatic and multistatic array architectures. The following theorem will be useful in characterizing the SBP of 1D imaging system.

\begin{theorem}
Let $\hat{\mathcal{T}}  = \mathcal{I}_{l} (\mathcal{T})$ denote the mapping operation, projecting set $\mathcal{T}$ onto the line $l$ in 2D space. Then, 
\begin{equation}
\mathcal{I}_{l}(\Tmono) = \mathcal{I}_{l}(\Tmulti), ~\forall l,
\end{equation}
for any point scatterer in the scene. Consequently, the spatial frequency bandwidth $B(x')$, is the same for monostatic and multistatic architectures. See Appendix \ref{app:proof} for the proof.
\end{theorem}

The rest of this section is devoted to analytic and numerical computation of SBP for different imaging geometries. We consider the nominal values for wavelength $\lambda = 5$mm (corresponding to $60$GHz temporal frequency), the size of the aperture $L_1 = 15$cm, the size of the scene $L_2 = 10$cm, and the distance between the aperture and the scene $D = 20$cm for the simulations in Section \ref{sec:SBP}, unless stated otherwise. 


\subsection{SBP for 1D parallel planes geometry}
\label{sec:SBP_parallel}
\begin{figure}[b]
\centering
\includegraphics[scale=0.4] {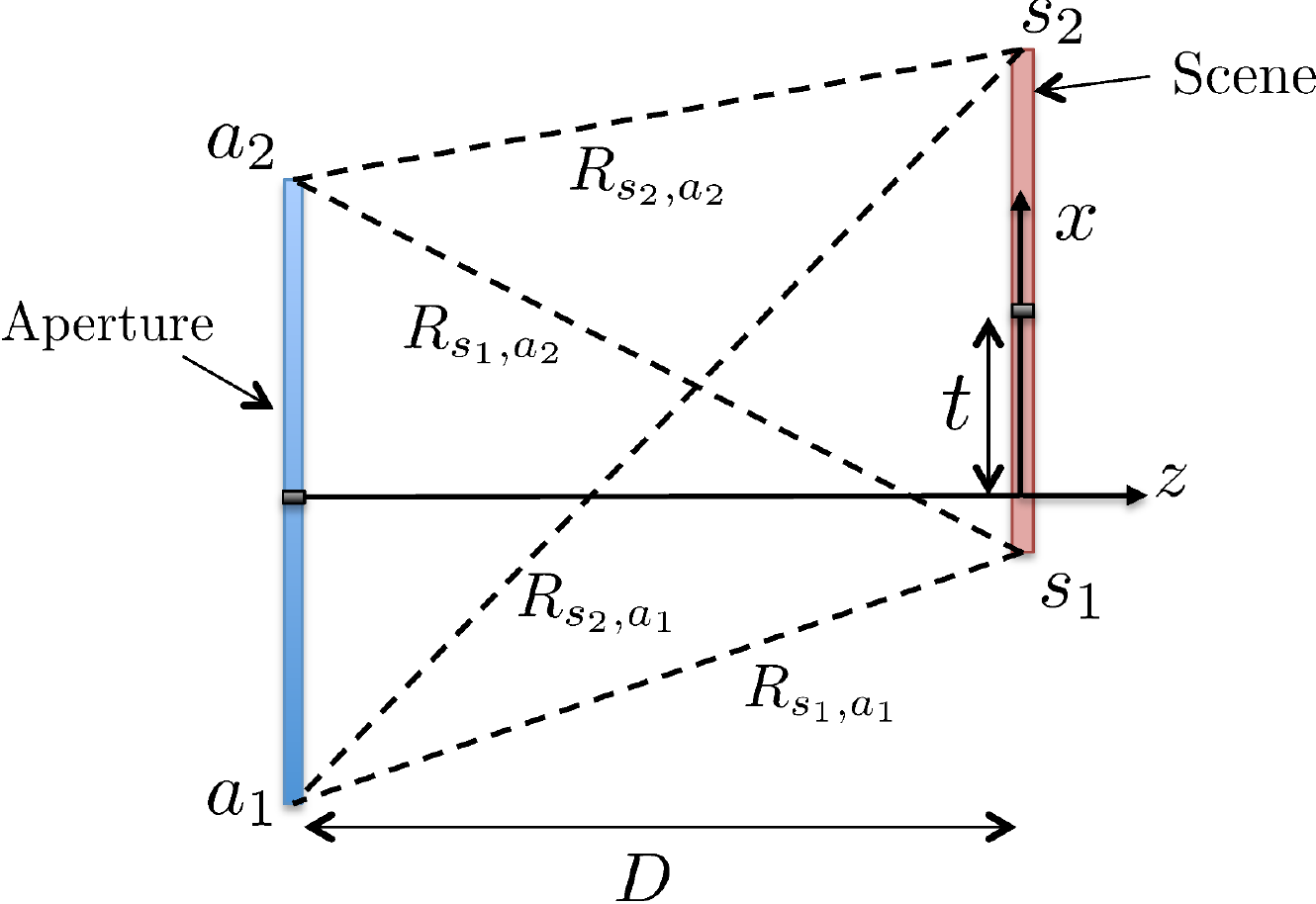}
\caption{Geometry G2: $1$D parallel planes propagation model, with an arbitrary scene translation $t$.}
\label{fig:parallel_geometry} 
\end{figure}
\begin{figure}[htbp]
\centering
\includegraphics[scale=0.13] {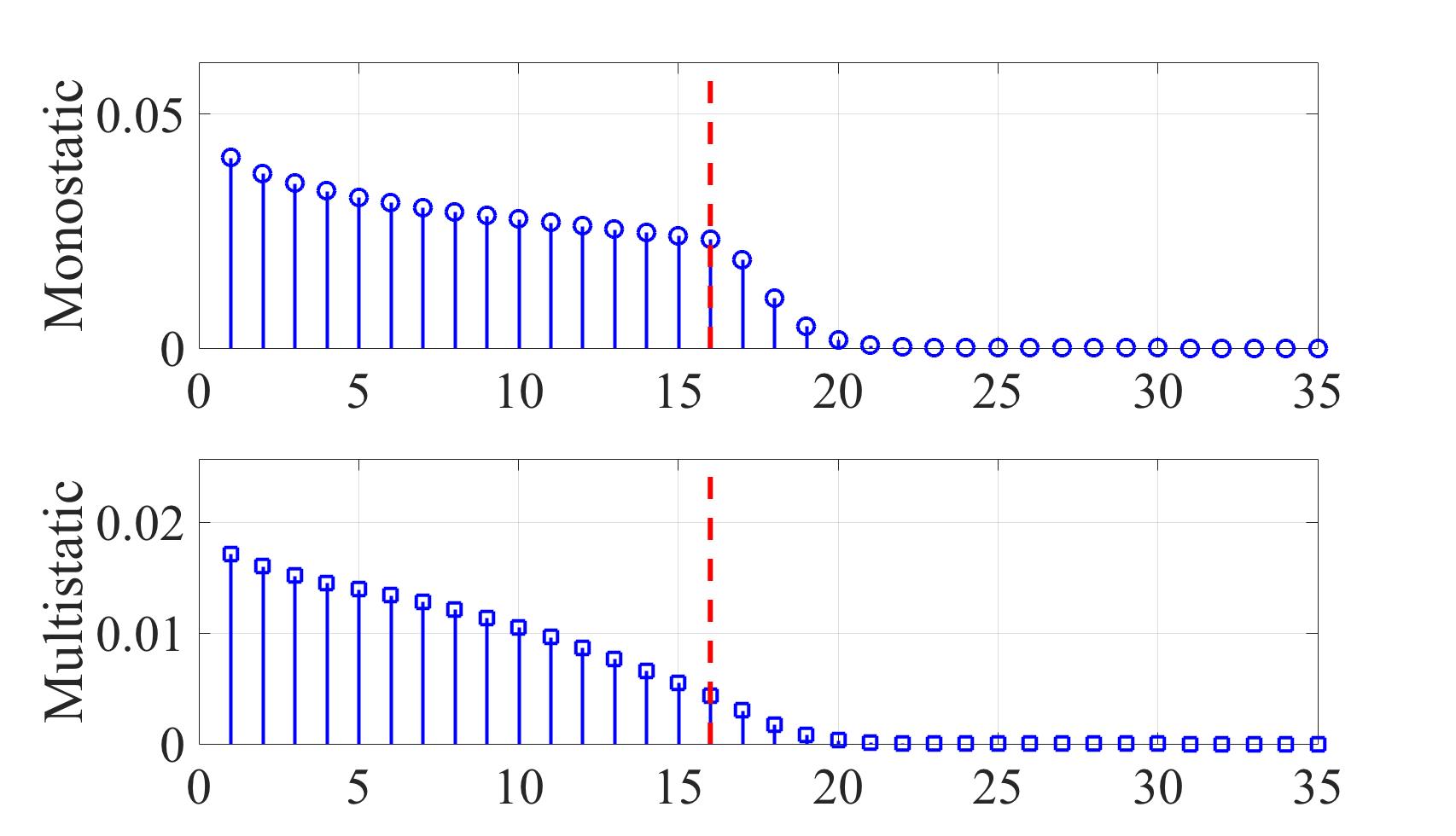}
  \caption{SVD analysis for parallel asymmetric geometry G2, with $t = 15$cm for (top) monostatic and (bottom) multistatic array of $N=200$ array elements. Note that for this geometry $\text{SBP}_{G2} \approx 16$, depicted by the dashed line. The estimate of the DoF based on the operator norm yields $\bar{\Sigma}_{sq} \approx 9.05$ and $7.69$, for monostatic and multistatic arrays, respectively, indicating that $\bar{\Sigma}_{sq}$ underestimates the DoF when the singular values do not follow a step-like behavior.}
  \label{fig:svd_trans_15} 
\end{figure}
\begin{figure}[htbp]
\centering
\includegraphics[scale=0.21] {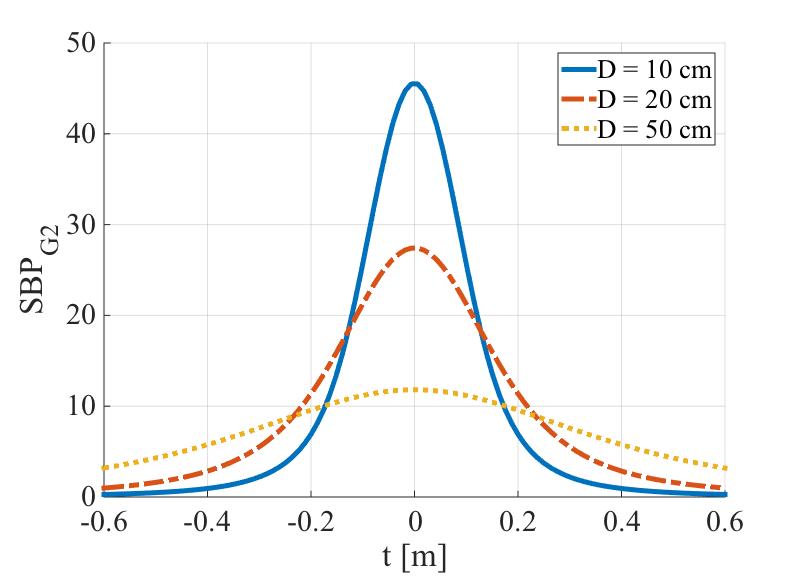}
  \caption{SBP computed for geometry G2 as a function of scene translation parameter $t$, with $L_1 = 15$ cm, $L_2 = 10$ cm, and $D\in\{10, 20,50\}$ cm. }
  \label{fig:G2_vs_t} 
\end{figure}

Consider the parallel planes geometry G2 with translation parameter $t$, shown in Figure \ref{fig:parallel_geometry}. For a point scatterer located at $x=x'$, we have
\begin{equation}
B(x') = {2\over \lambda}\left(\sin(\beta(x')) - \sin(\alpha(x'))\right),
\end{equation}
where, 
\begin{eqnarray}
\sin(\beta(x')) = {(x' - a_1) \over \sqrt{D^2 + (x' - a_1)^2}}\\
\sin(\alpha(x')) = {(x' - a_2) \over \sqrt{D^2 + (x' - a_2)^2}}.
\end{eqnarray}
For a small segment of the scene of length $dx'$, the SBP is approximately given by $B(x')dx'$. Therefore, the total SBP of the scene is calculated by the integral
\begin{equation}
\text{SBP} = \int_{\text{scene}} B(x')dx'.
\end{equation} 
Assuming the scene reflectivity function $\bgamma(x') = 0$ for $x'\notin[s_1,s_2]$, and the aperture spanning an interval $[a_1,a_2]$ (as shown in Figure \ref{fig:parallel_geometry}), the SBP is calculated as
\begin{eqnarray}
\text{SBP}_{G2}\!\! &=& \int_{s_1}^{s_2} B(x')dx' \notag\\
&=& \int_{s_1}^{s_2} {2\over \lambda} (\sin(\beta(x')) - \sin(\alpha(x')))dx' \notag\\
 &=& \!\!\!\! {2\over \lambda}\Big((R_{s_2, a_1} - R_{s_2, a_2}) + (R_{s_1, a_2} - R_{s_1, a_1})\! \Big) \notag\\
\end{eqnarray}
where $R_{i,j}$ denotes the distances between points $i$ and $j$. Figure \ref{fig:svd_trans_15} shows the singular values (normalized by a scalar factor) of the discretized integral operator for $t = 15$cm for monostatic and multistatic architectures. We see that $\text{SBP}_{G2}$, shown by the dashed lines, accurately predicts the critical point after which the singular values drop quickly to zero. We have  conducted simulations for various values of parameters $L_1, L_2, D$ and $t$, to verify the accuracy of $\text{SBP}_{G2}$ in identifying the DoF for geometry G2. Figure \ref{fig:G2_vs_t} shows the variation of $\text{SBP}_{G2}$ as a function of the translation parameter $t$, at various distances.  We see that the maximum $\text{SBP}_{G2}$ is obtained at $t=0$ for all $D$, with the sensitivity of SBP to $t$ being inversely related to $D$.

\begin{figure}[htbp]
\centering 
\subfigure[]{
\includegraphics[scale=0.18]{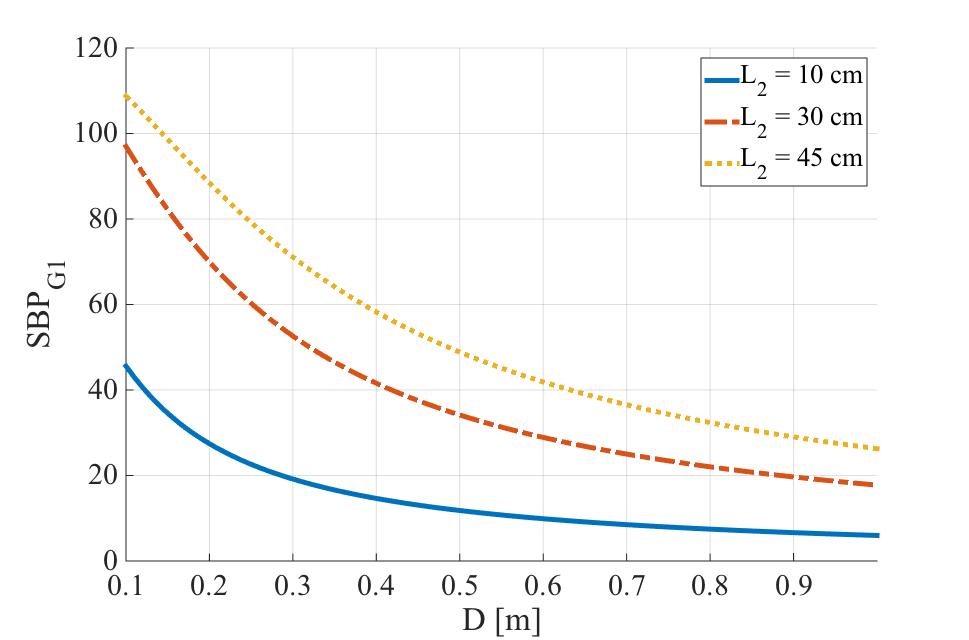}
}
\subfigure[]{
\includegraphics[scale=0.18]{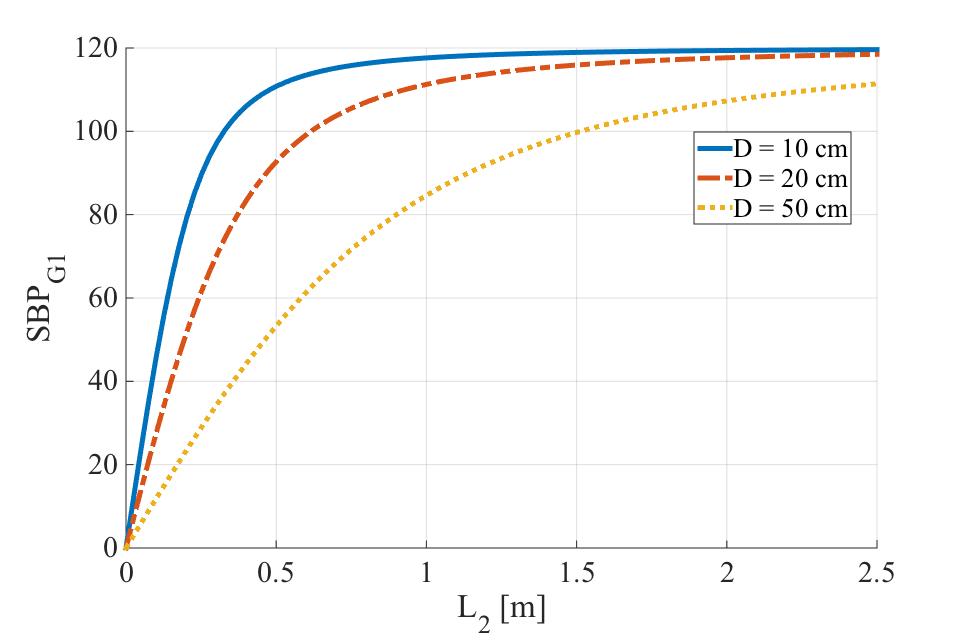}
}
\caption{SBP computed for geometry G1 with $L_1 = 15$ cm fixed, as a function of (a) distance $D$, and (b) scene extent $L_2$.}
\label{fig:G1_vs_D} 
\end{figure}

For the special case of symmetric and parallel geometry G1 (shown in Figure \ref{fig:geometry_G1}), we have $[a_1, a_2] = [-L_1/2, L_1/2]$ and $[s_1, s_2] = [-L_2/2, L_2/2]$, 
hence the space-bandwidth product is given by
\begin{eqnarray}
&\text{SBP}_{G1}& = {4\over \lambda} (R_{s_2, a_1} - R_{s_2, a_2}) \notag\\
&= & \!\!\!\!\!\!\!\!  {4D\over \lambda}\left(\sqrt{1 + \left({L_1+L_2\over2D}\right)^2} - \sqrt{1 + \left({L_1-L_2\over2D}\right)^2}\right).\notag\\
\label{eq:SBP_G1}
\end{eqnarray}
\begin{figure}[t]
\centering
\includegraphics[scale=0.13] {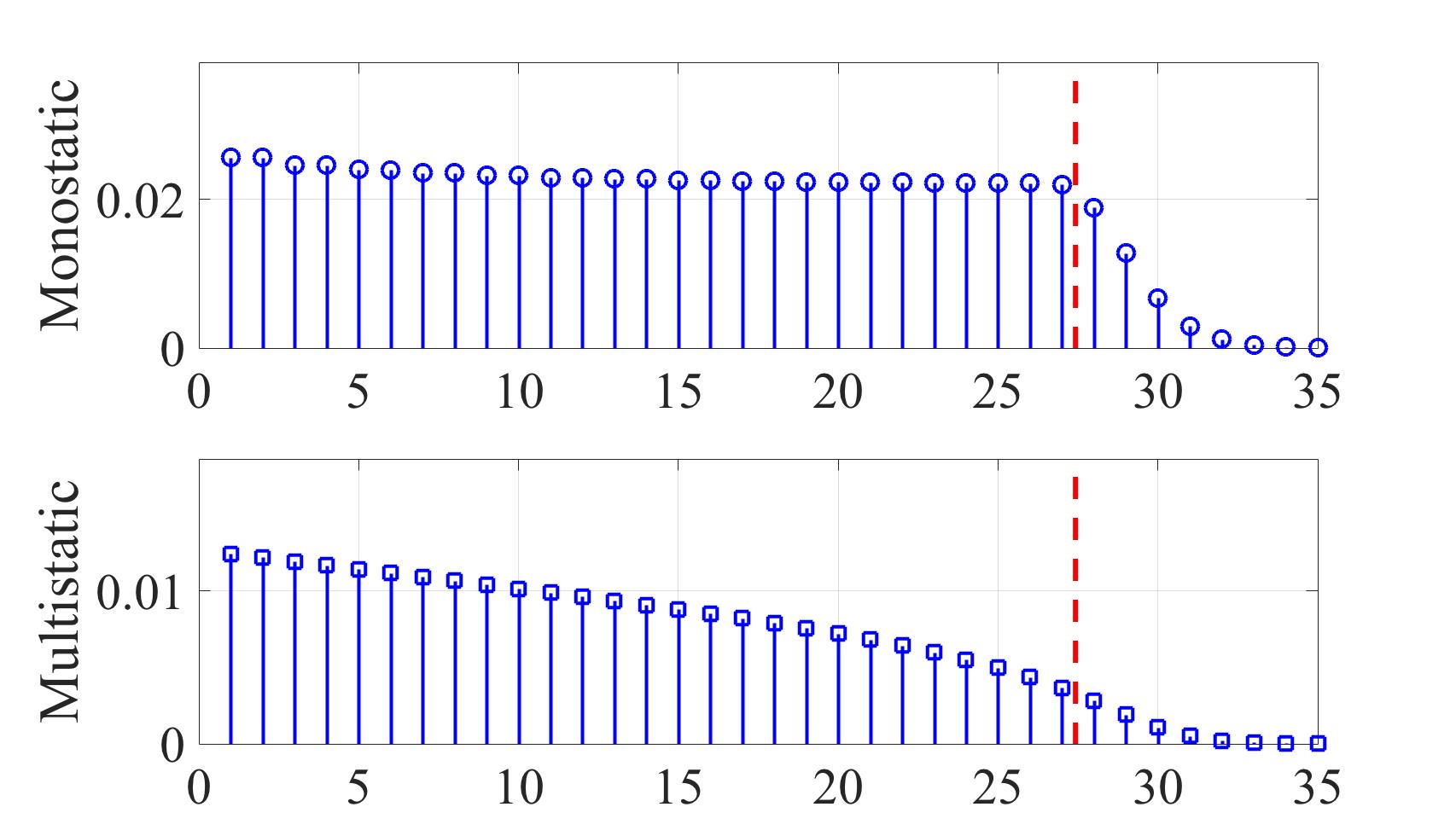} 
  \caption{SVD analysis for nominal symmetric geometry G1, for (top) monostatic and (bottom) multistatic array of $N=200$ array elements. Note that for this geometry $\text{SBP}_{G1} \approx 27.4$, depicted by the dashed line.  The estimate of the DoF based on the operator norm yields $\bar{\Sigma}_{sq} \approx 22.88$ and $14.63$, for monostatic and multistatic arrays, respectively.}
  \label{fig:svd_trans_0} 
\end{figure}
For unbounded aperture $\lim_{L_1\to\infty} \text{SBP}_{G1} = {4L_2\over \lambda}$. Similarly, for unbounded scene $\lim_{L_2\to\infty} \text{SBP}_{G1} = {4L_1\over \lambda}$.
Thus, $\text{SBP}_{G1}$ does not increase indefinitely with increasing aperture or scene size.  We note that our $\text{SBP}_{G1}$ calculations are consistent with the heuristics reported in \cite{Pierri_heuristic}, and the explicit derivations in \cite{Solimene_Strip_Sources} for bounded and unbounded observation domains. 
As depicted in Figure \ref{fig:svd_trans_0}, $\text{SBP}_{G1}$ can accurately predict the number of DoF for the nominal geometry G1. Figure \ref{fig:G1_vs_D} summarizes the behavior of $\text{SBP}_{G1}$ as a function of $D$ and $L_2$. In particular Figure \ref{fig:G1_vs_D}-b shows that $\text{SBP}_{G1}$ reaches a plateau as $L_2$ is increased, with the corresponding upper bound being independent of $D$.

\subsection{SBP for 1D rotated planes geometry}
\label{sec:SBP_rotated}

\begin{figure}[htbp]
\centering
\includegraphics[scale=0.4] {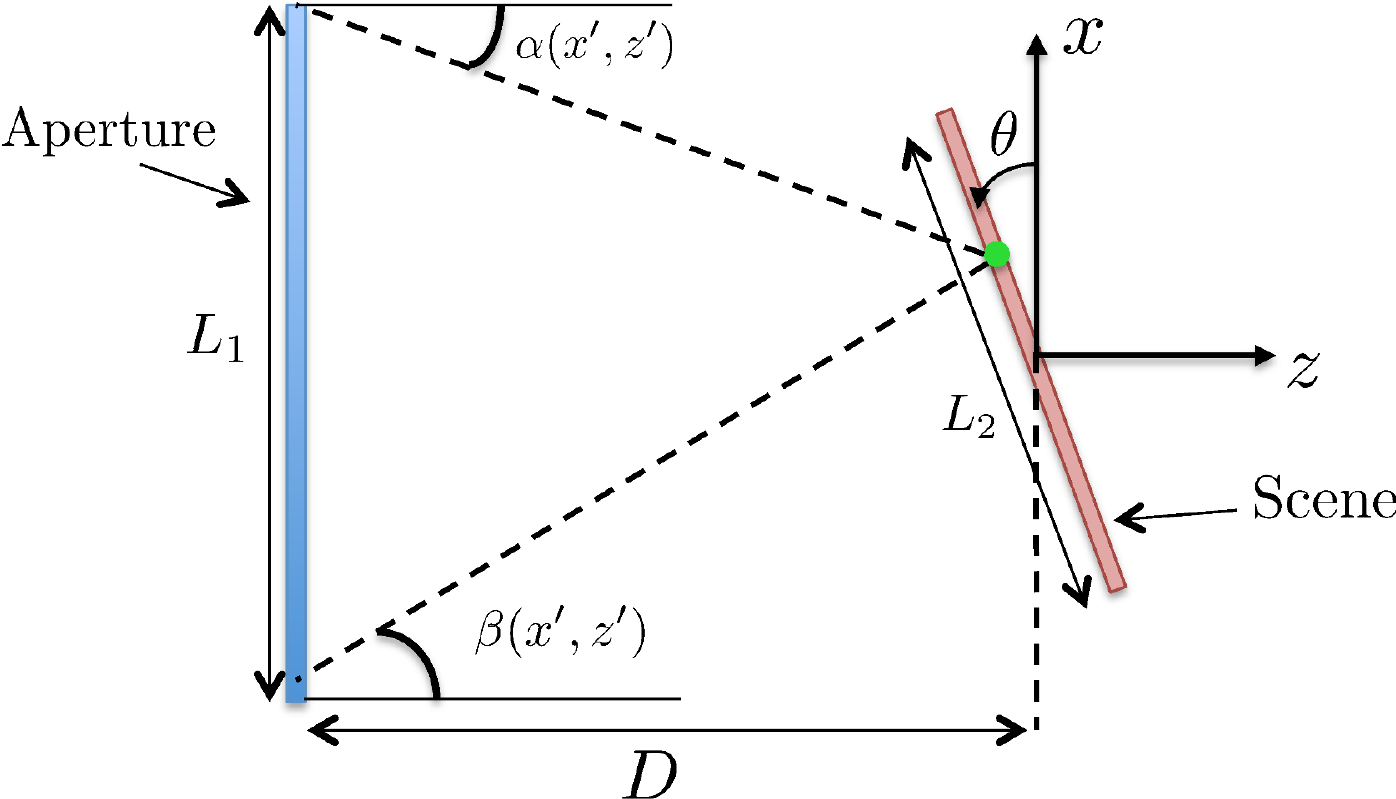}
  \caption{Geometry G3: $1$D rotated planes propagation model.}
  \label{fig:G3} 
\end{figure}
\begin{figure}[htbp]
\centering
\includegraphics[scale=0.4] {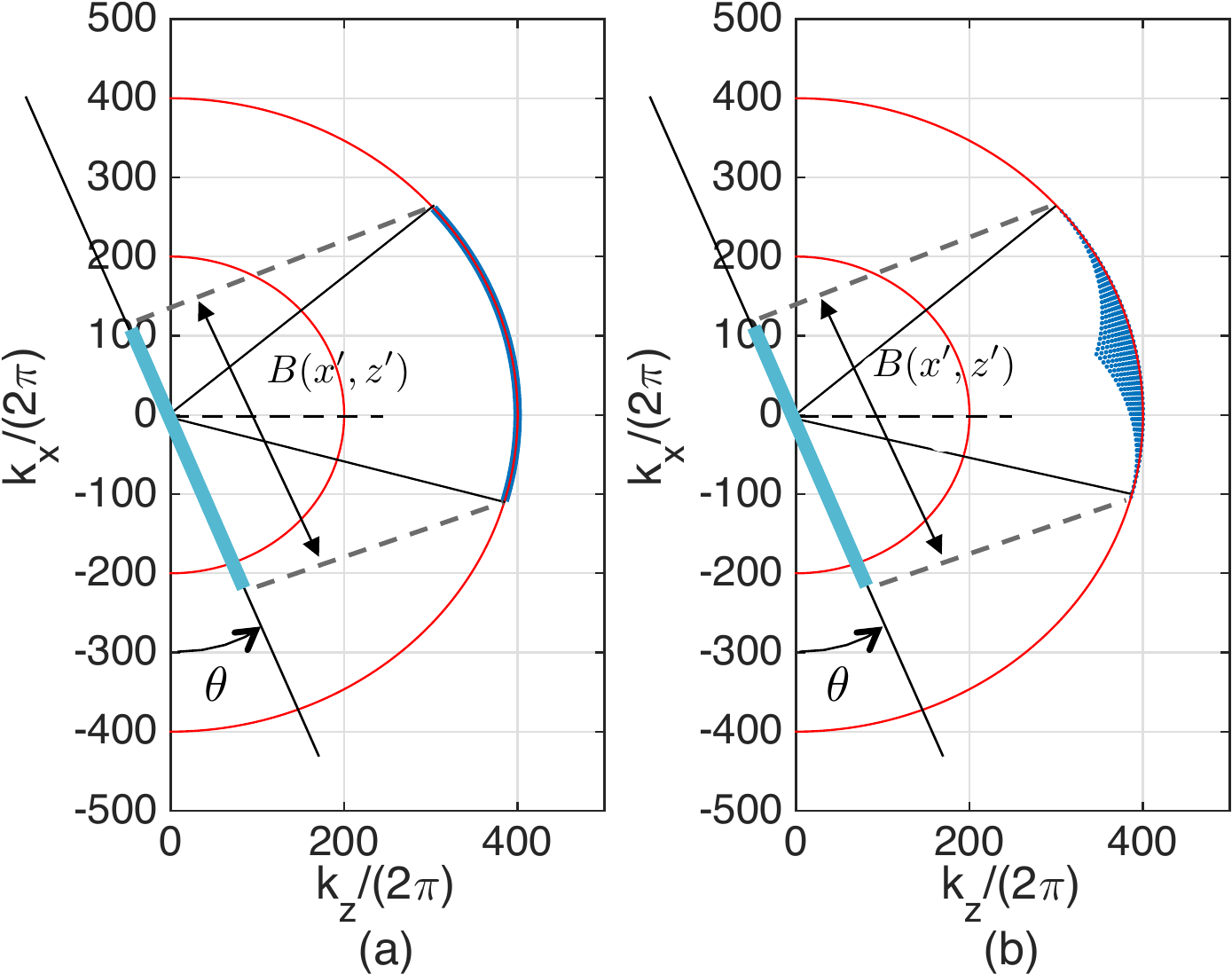}
  \caption{Spatial-frequency bandwidth corresponding to the point scatterer located at $(x',z')$, for (a) monostatic and (b) multistatic array of infinitely many TRx elements.}
  \label{fig:bandwidth_rotated} 
\end{figure}
\begin{figure}[htbp]
\centering
\includegraphics[scale=0.13] {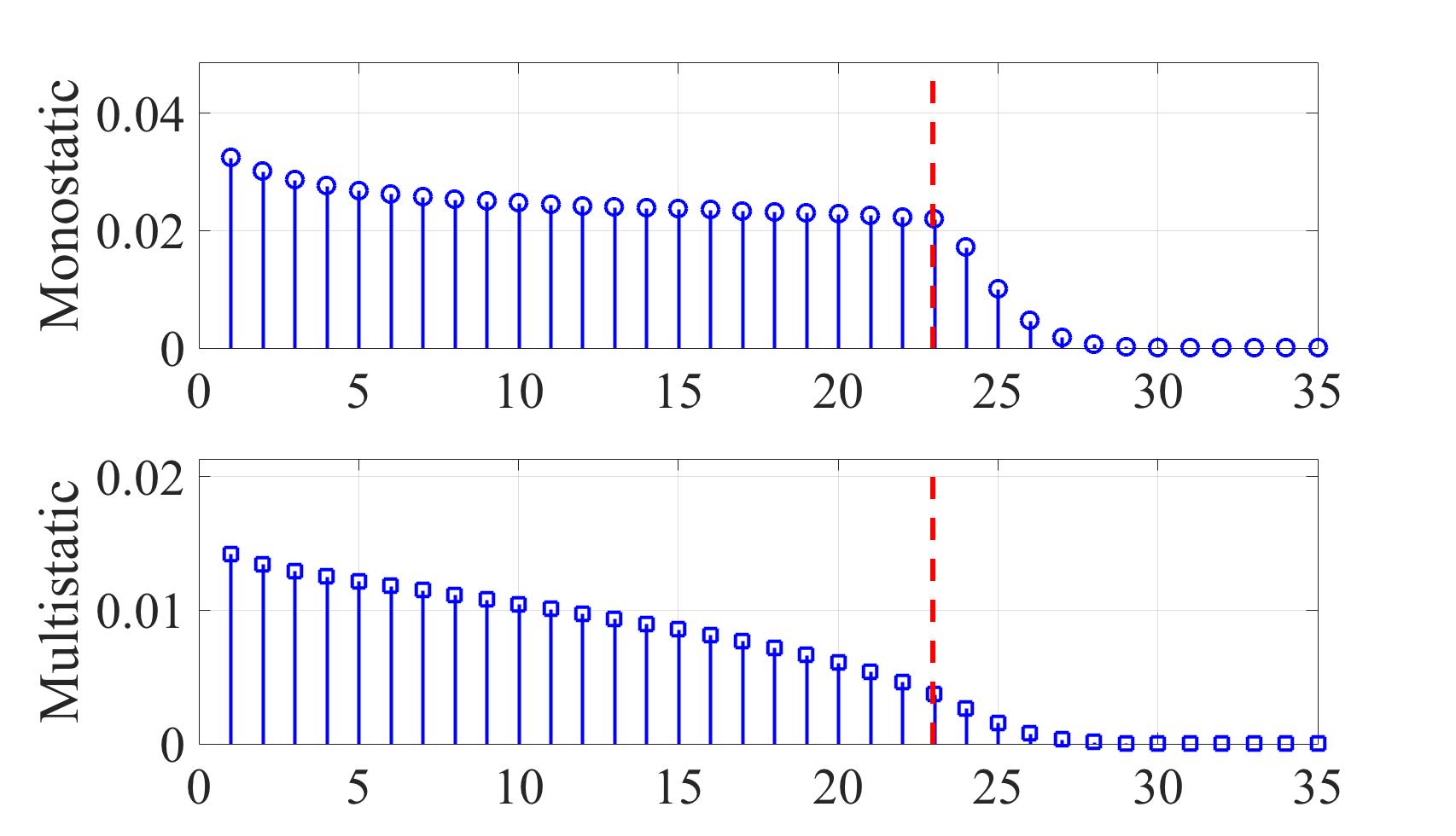}
  \caption{SVD analysis for rotated geometry G3, with $\theta = 35^{\circ}$ for (top) monostatic and (bottom) multistatic array of $N=200$ array elements. Note that for this geometry $\text{SBP}_{G3} \approx 23$, depicted by the dashed line. The estimate of the DoF based on the operator norm yields $\bar{\Sigma}_{sq} \approx 14.24$ and $11.14$, for monostatic and multistatic arrays, respectively.}
  \label{fig:svd_rot_35} 
\end{figure}

Consider the geometry G3 depicted in Figure \ref{fig:G3}, where the scene creates an angle $\theta$ with the $x$ coordinate. In this scenario, the reflectivity function $\bgamma(x',z')$ is restricted to the line \mbox{$x' = \rho z'$}, where $\rho = {-1\over \tan(\theta)}$, i.e., $\bgamma(x',z') = 0$ for all $x' \neq \rho z'$. Rewriting equation (\ref{eq:sample_gamma}) under this constraint leads to
\begin{eqnarray}
\label{eq:spectrum_G3}
&&S(k_{x_{tx}} , k_{x_{rx}}) =  e^{j(k_{z_{tx}}+k_{z_{rx}})z_{a}} \times \notag\\
&&\int\limits_{z'} \bgamma(\rho z', z') e^{-j(k_{x_{tx}} + k_{x_{rx}})\rho z'} e^{-j(k_{z_{tx}} + k_{z_{rx}}) z'} dz'\notag\\
&&= \int\limits_{z'} \bgamma(\rho z', z') e^{-j (\rho k_x + k_z) z'} dz'.
\end{eqnarray}
The integral kernel in (\ref{eq:spectrum_G3})  depends on $\rho k_{x} + k_z$, i.e., any pair of Tx/Rx elements that lead to the same value of $\rho k_{x} + k_{z}$, deliver the same information about the reflectivity function. In order to avoid redundancy in the acquired information through the imaging system, we need to project the sampled points in the spectrum of the scene onto the line $k_{x} = \rho k_{z}$, i.e., the line that crosses the origin and creates an angle $\theta$ with the $k_{x}$ coordinate, as shown in Figure \ref{fig:bandwidth_rotated}.  The space-bandwidth product for this geometry is given by 
\begin{equation}
\text{SBP}_{G3} = \int_{\text{scene}} B(x',z') d\mu(x',z').
\end{equation}
where $\mu(\cdot,\cdot)$ is the standard Lebesgue measure on $\mathcal{A}$. As shown in Theorem 1, for any angle $\theta$, the spatial frequency bandwidth $B(x', z')$, and consequently $\text{SBP}_{G3}$, is the same for monostatic and multistatic arrays. This can be verified through numerical SVD computations shown in Figure \ref{fig:svd_rot_35} for $\theta = 35^{\circ}$. 

\begin{figure}[htbp]
\centering
\includegraphics[scale=0.23] {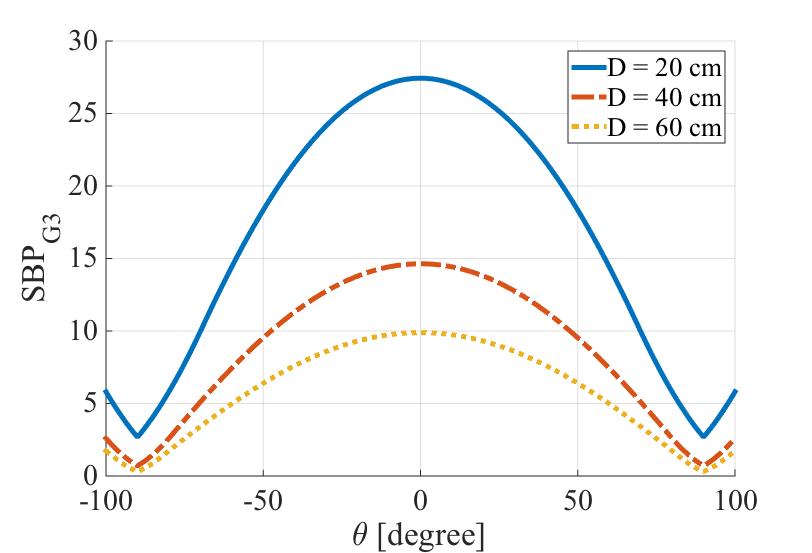}
  \caption{SBP computed for geometry G3 as a function of scene rotation $\theta$, with $L_1 = 15$ cm, $L_2 = 10$ cm, and $D\in\{20, 40, 60\}$ cm.}
  \label{fig:SBP_G3_theta} 
\end{figure}

Figure \ref{fig:SBP_G3_theta} shows the result of numerical computation of $\text{SBP}_{G3}$ for different rotation angles and at different ranges. The space-bandwidth product is maximized for $\theta = 0$, which corresponds to parallel symmetric geometry G1. Increasing $\theta$ leads to a decrease in $\text{SBP}_{G3}$, till a global minimum is achieved at $\theta = 90^{\circ}$ (i.e., when the planes are orthogonal to each other). Note that, even for the orthogonal planes geometry, $\text{SBP}_{G3}$ is bounded away from zero, with its value increasing as we decrease the distance $D$. This is indeed the reason behind the improvement in range resolution for continuous-wave imaging systems at short range \cite{Ahmed_thesis}.

\subsection{SBP for 1D rotated and translated planes geometry}
\begin{figure}[htbp]
\centering
\includegraphics[scale=0.45] {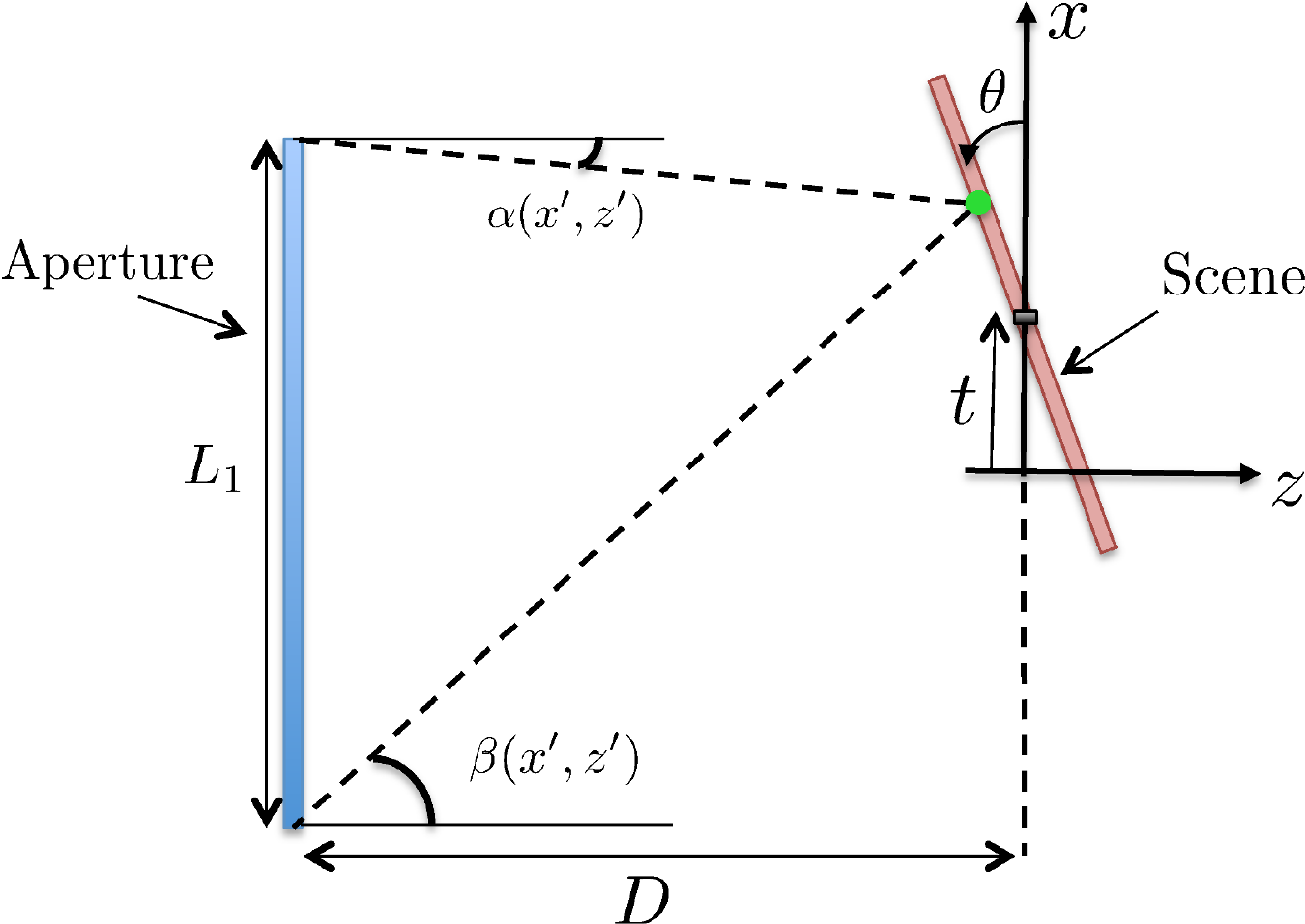}
  \caption{Geometry G4: $1$D rotated and translated planes propagation model.}
  \label{fig:G4} 
\end{figure}

\begin{figure}[htbp]
\centering
\includegraphics[scale=0.13] {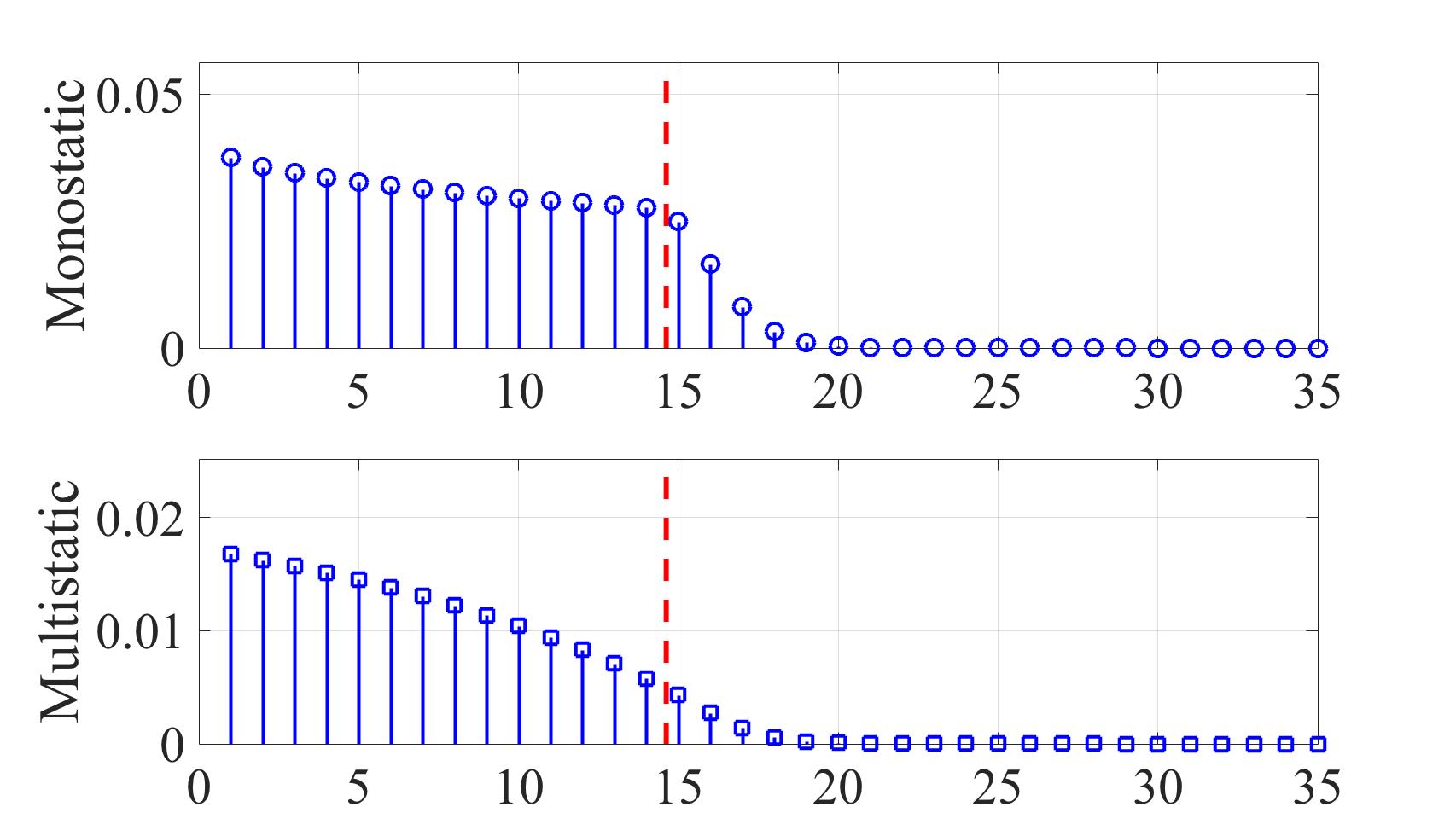}
  \caption{SVD analysis for geometry G4 with  $t = 20$ cm, and $\theta = 55^{\circ}$, for (top) monostatic and (bottom) multistatic array of $N=200$ array elements. Note that for this geometry $\text{SBP}_{G4} \approx 14.6$, depicted by the dashed line. The estimate of the DoF based on the operator norm yields $\bar{\Sigma}_{sq} \approx 10.62$ and $8$, for monostatic and multistatic arrays, respectively.}
  \label{fig:svd_G4} 
\end{figure}

\begin{figure}[htbp]
\centering
\includegraphics[scale=0.2] {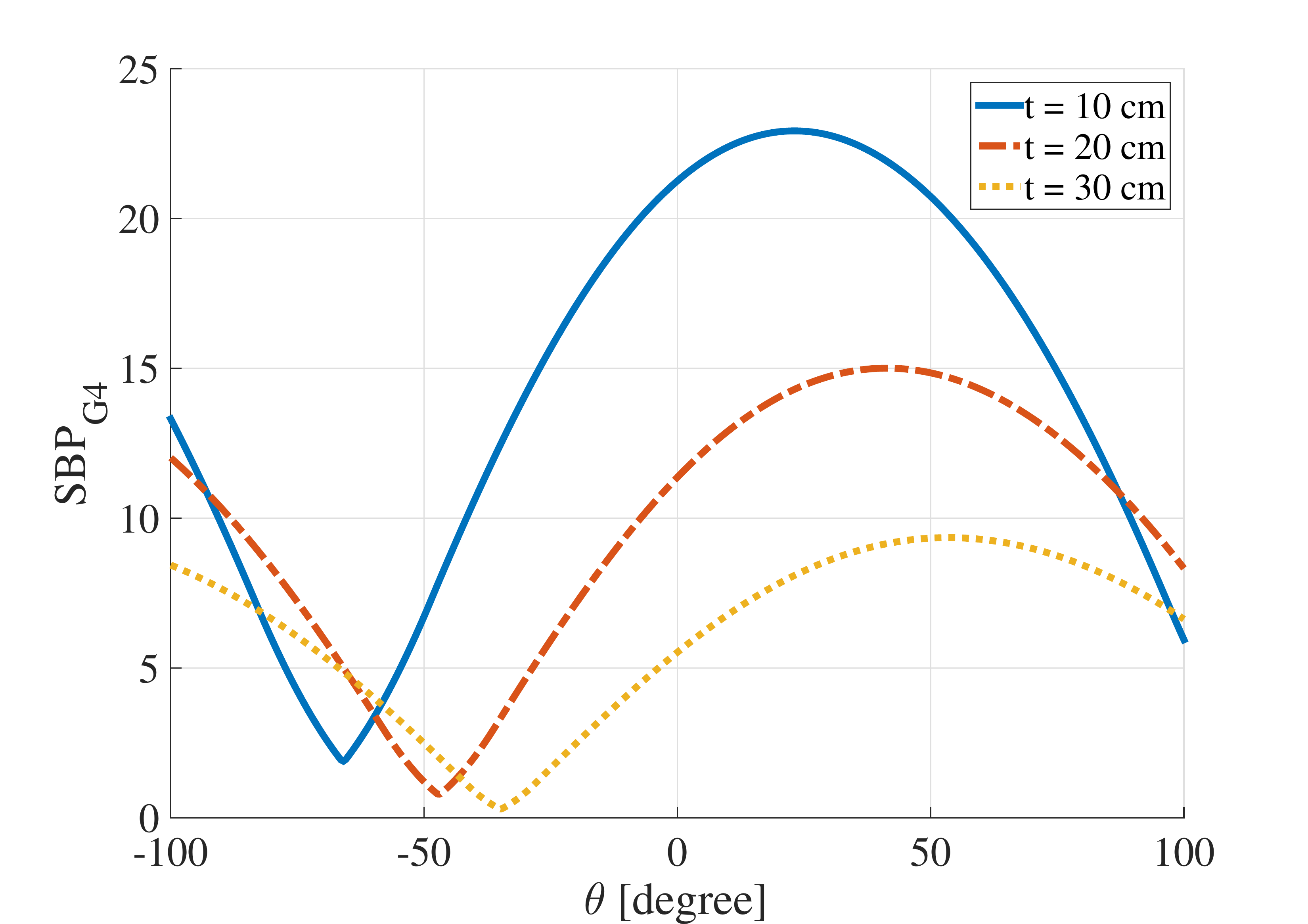}
  \caption{SBP computed for geometry G4, with $L_1 = 15$cm, $L_2 = 10$cm, $D = 20$cm, and $t\in\{10, 20, 30\}$ cm.}
  \label{fig:G4_vs_theta_t} 
\end{figure}

\begin{figure}[htbp]
\centering
\includegraphics[scale=0.2] {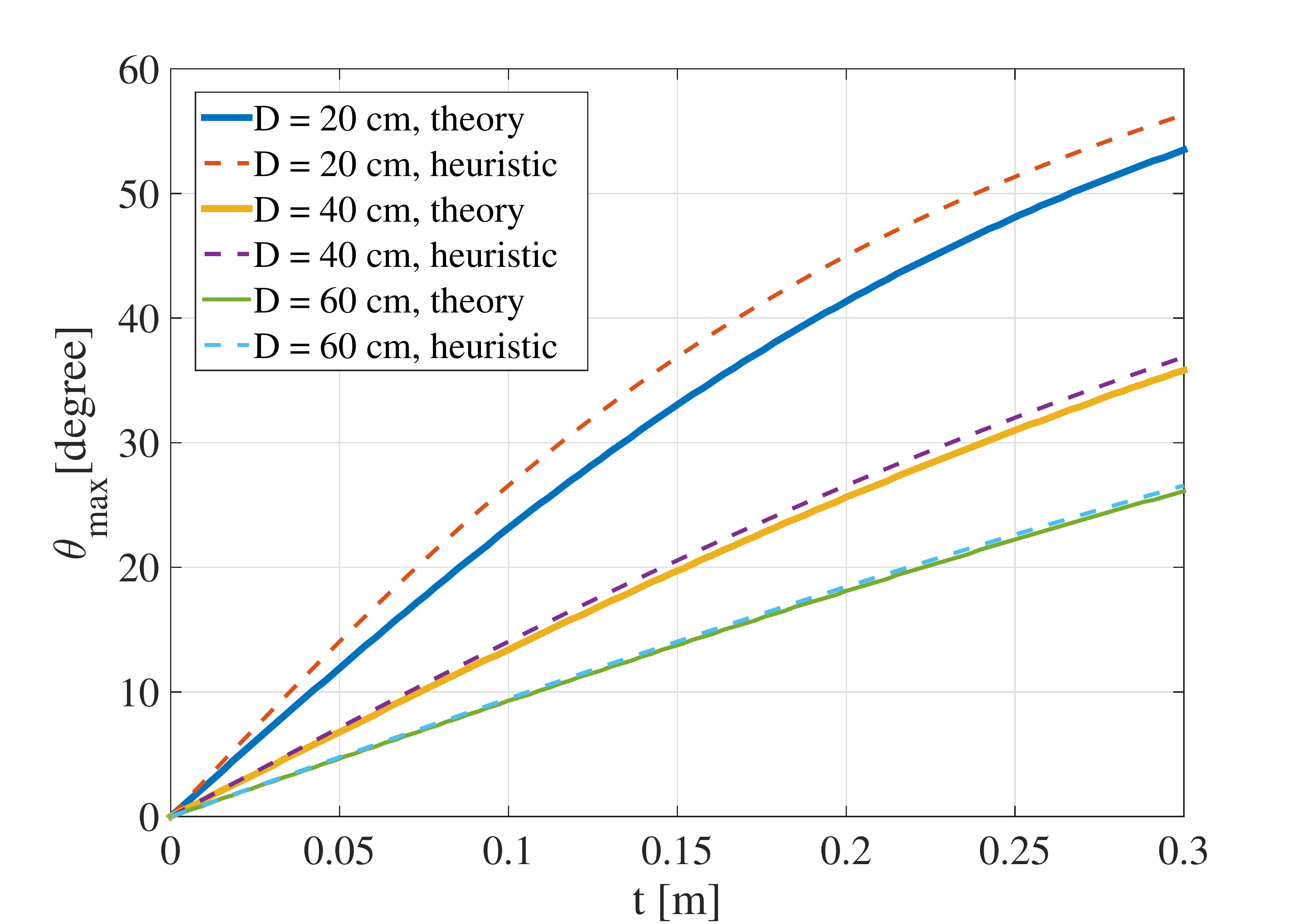}
  \caption{Rotation angle corresponding to the maximum $\text{SBP}_{G4}$ as a function of translation parameter $t$, computed for $D\in\{20, 40, 60\}$ cm. The dashed and solid curves correspond to $\theta_{heu}$ and true $\theta_{max}$, respectively.}
  \label{fig:G4_theta_max} 
\end{figure}

We now consider the geometry depicted in Figure \ref{fig:G4}, where the scene is rotated and translated simultaneously. The scene reflectivity function is restricted to the line $x' = \rho z' + t$, where $\rho = {-1\over\tan(\theta)}$. Following similar arguments as in Subsection \ref{sec:SBP_rotated}, one can show that the spatial frequency bandwidth for any point scatterer in the scene is evaluated by projecting the sampled points in the scene spectrum onto the line $k_x = \rho k_z$ (as depicted in Figure \ref{fig:bandwidth_rotated}). We compute $\text{SBP}_{G4}$ numerically for different realizations of the scene parameters. Figure \ref{fig:svd_G4} shows the singular values and the corresponding $\text{SBP}_{G4}$ for $\theta = 55^{\circ}$ and $t = 20$cm. The variation of $\text{SBP}_{G4}$ as a function of $\theta$ for multiple values of $t$ is depicted in Figure \ref{fig:G4_vs_theta_t}. It is interesting to determine $\theta_{max}(t)$, i.e., the rotation angle that maximizes $\text{SBP}_{G4}$ for a given $t$. One heuristic approach is based on choosing $\theta$ such that the scene is orthogonal to the line that connects the midpoints of the aperture and the scene, i.e., $\theta_{heu}(t) = \sin^{-1}(t/\sqrt{t^2 + D^2})$. Figure \ref{fig:G4_theta_max} shows that, at short distances, $\theta_{heu}$ overestimates $\theta_{max}$ for $t>0$, with the difference vanishing as $D$ is increased.

\section{The Fresnel Regime}
\label{sec:Fresnel}
In this section, we study monostatic and multistatic arrays under the Fresnel approximation \cite{Lee_book}. That is, we use a first order Taylor approximation for computing path lengths in (\ref{eq:born}), assuming $D \gg L_1, L_2$. For the geometry G1 shown in Figure \ref{fig:geometry_G1}, this approximation yields
\begin{equation}
R(x,x';z' = 0) = \sqrt{(x-x')^ + D^2} \approx D + {(x-x')^2\over 2D}.
\end{equation}
Therefore, the Fresnel diffraction integral is given by
\begin{eqnarray}
s(x_{tx}, x_{rx}) = \int\limits_{x'} \bgamma(x') e^{-jkR(x_{tx}, x';0)} e^{-jkR(x_{rx},x';0)} dx'\notag\\
\approx e^{-j2kD} \int\limits_{x'} \bgamma(x') e^{-j {k\over 2D}(x_{tx} - x')^2} e^{-j {k\over 2D}(x_{rx} - x')^2} dx'. 
\label{eq:fresnel}
\end{eqnarray}
We investigate the implications of the the Fresnel approximation for array design, and illustrate the connections between different array architectures in the Fresnel regime. 

\subsection{Monostatic array in the Fresnel regime}
\label{sec:mono_Fresnel}
\begin{figure}[htbp]
\centering 
\subfigure[]{
\includegraphics[scale=0.13]{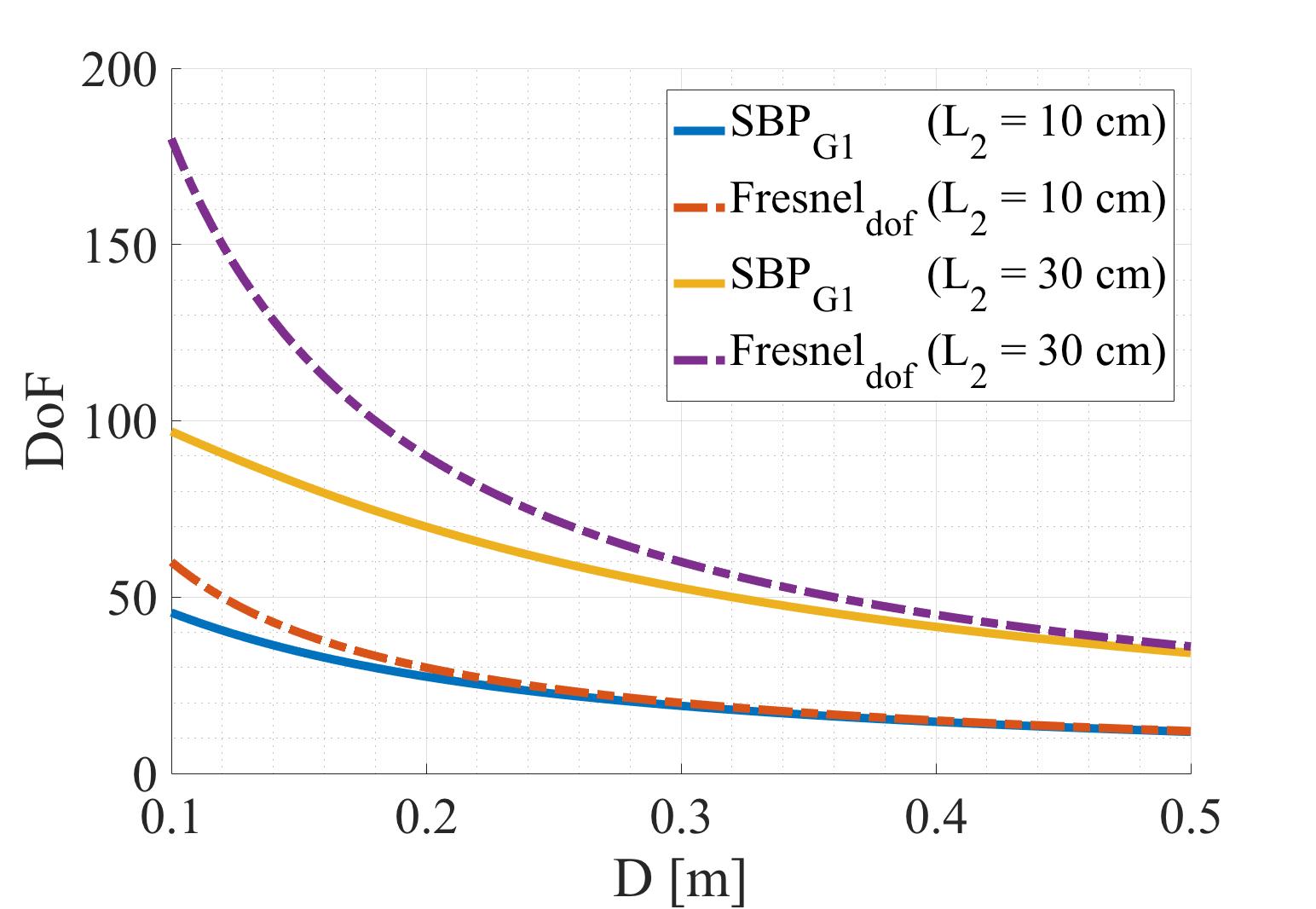}
}
\subfigure[]{
\includegraphics[scale=0.13] {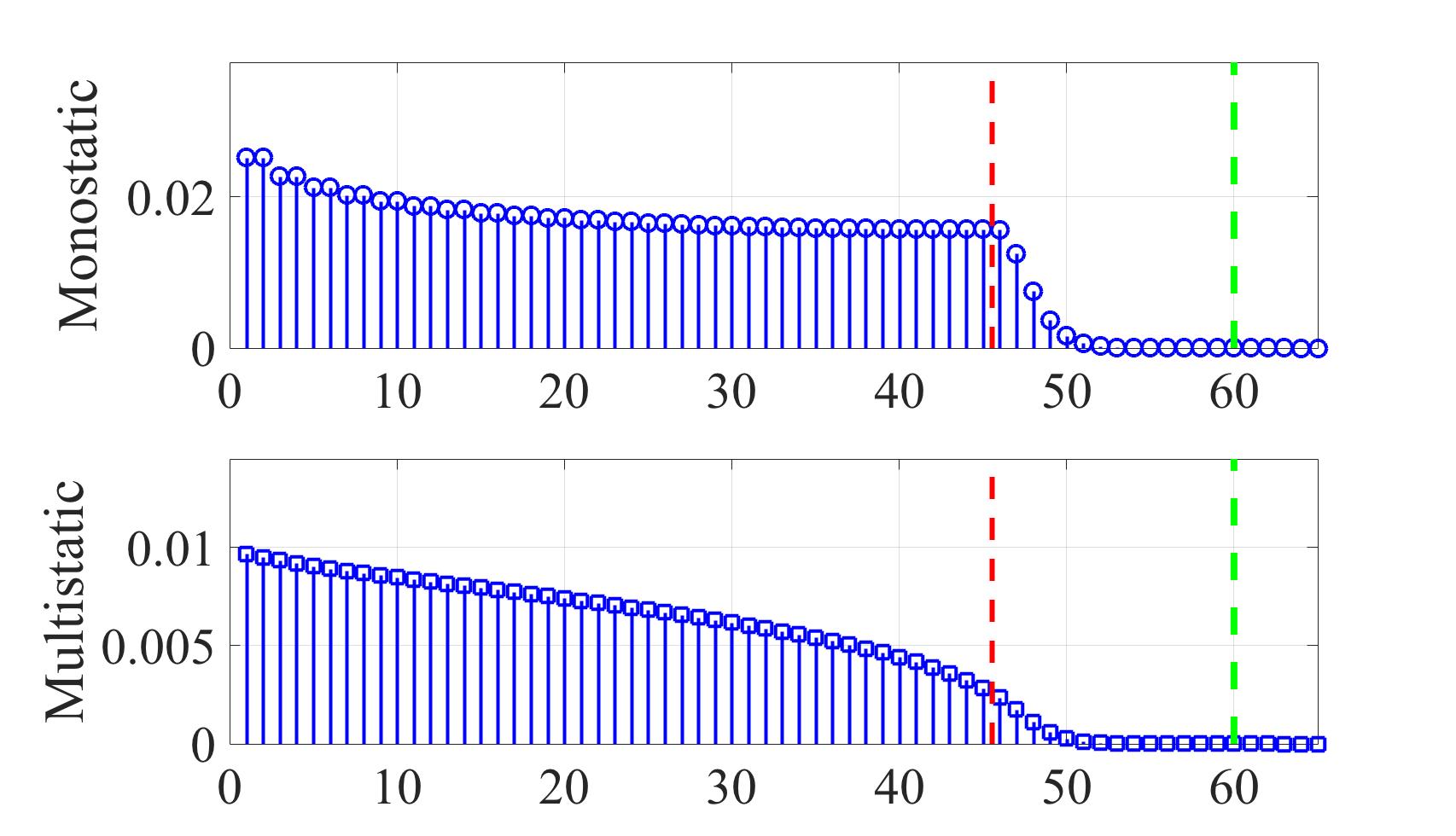}
}
\caption{(a) DoF predicted by $\text{SBP}_{G1}$ and the the Fresnel approximation, computed for geometry G1 with $L_1 = 15$cm fixed, as a function of distance $D$, and (b) SVD analysis for nominal symmetric geometry G1, with $L_1 = 15$cm, $L_2 = 10$cm, and $D = 10$cm, for (top) monostatic and (bottom) multistatic array of $N=200$ array elements. Note that for this geometry $\text{SBP}_{G1} \approx 45.6$, and $\text{Fresnel}_{dof} = 60$, depicted by the dashed red and green lines, respectively. The operator norm-based estimator $\bar{\Sigma}_{sq}$ again underestimates the DoF, yielding $23.6$ and $24.08$, for monostatic and multistatic arrays, respectively.}
\label{fig:SBP_vs_Fresnel}  
\end{figure}

Since $x_{tx} = x_{rx}$ for a monostatic architecture, equation (\ref{eq:fresnel}) reduces to
\begin{eqnarray}
&&s(x_{tx}, x_{rx})\approx  e^{-j2kD} \int\limits_{x'} \bgamma(x') e^{-j {k\over D}(x_{tx} - x')^2} dx' \notag\\
&&= e^{-j2kD} e^{-j{k\over D}x_{tx}^{2}} \int\limits_{x'} \bgamma(x') e^{-j{k\over D}x'^{2}} e^{j({2kx_{tx}\over D})x'} dx'\notag\\
&&=  e^{-j2kD} e^{-j{k\over D}x_{tx}^{2}} \text{FT}_{1D} \left\{\bgamma(x')e^{-j{k\over D}x'^{2}}\right\}_{f\triangleq {kx_{tx}\over \pi D}},
\label{eq:mono_Fresnel}
\end{eqnarray}
where we have used a change of variable $f\triangleq {kx_{tx}\over \pi D}$ in computing the 1D Fourier transform. The quadratic-phase terms (i.e., $e^{-j{k\over D} x'^{2}}$ and $e^{-j{k\over D} x_{tx}^{2}}$) in (\ref{eq:mono_Fresnel}) are known as {\it Fresnel phase masks} \cite{Lee_book}. Multiplying the reflectivity function by such a mask does not lead to any information loss, since its effect can be inverted using the complex conjugate mask. Therefore, the only information bottleneck in equation (\ref{eq:mono_Fresnel}) is due to the Fourier Trasform operation. As mentioned in Section \ref{sec:introduction}, the Fourier kernel has been studied in detail by Slepian {\it et al} in the context of time-limited and band-limited functions \cite{Slepian}. It has been shown that the eigenfunctions of this integral equation are PSWFs, and the corresponding eigenvalues have the interesting property that they remain approximately equal until a critical transition point, after which they rapidly decay to zero. This transition point for the class of time- and band-limited signals is determined by the time-bandwidth product \cite{PSWF_three}. The equivalent of time-bandwidth product in (\ref{eq:mono_Fresnel}) corresponds to, 
\begin{equation}
\text{Fresnel}_{dof} = \Delta x' \Delta f = L_2 ({k\Delta x_{tx}\over \pi D}) = {2L_1L_2\over \lambda D},
\label{eq:DoF_Shannon}
\end{equation}
where $\Delta x' = L_2$ and $\Delta x_{tx} = L_1$ are identified  by the scene and aperture extent, respectively. Figure \ref{fig:SBP_vs_Fresnel}-a shows the DoF predicted by $\text{Fresnel}_{dof}$ compared to $\text{SBP}_{G1}$ for geometry G1. We see that when $D \gg L_1, L_2$ is not satisfied, $\text{Fresnel}_{dof}$ deviates from  $\text{SBP}_{G1}$. As an example, Figure \ref{fig:SBP_vs_Fresnel}-b depicts the singular values of the system for a specific realization of the geometry G1, along with the DoF predictions by  $\text{Fresnel}_{dof}$ and $\text{SBP}_{G1}$. We see that $\text{Fresnel}_{dof}$ significantly overestimates the available DoF in this scenario.

In Section \ref{sec:SBP_parallel}, we derived a closed-form expression for the space-bandwidth product for the symmetric parallel planes geometry G1, without any assumption on the distance of the scene from the aperture. In the special case of $D\gg L_1, L_2$, we can use the the Fresnel approximation to evaluate $\text{SBP}_{G1}$ as
\begin{eqnarray}
&\text{SBP}_{G1}& = {4\over \lambda} (R_{s_2, a_1} - R_{s_2, a_2}) \notag\\
&\approx& \!\!\!\!\!\!\!\!  {4D\over \lambda}\left(1 + {1\over 2}\left({L_1+L_2\over2D}\right)^2 - 1 - {1\over 2}\left({L_1-L_2\over2D}\right)^2 \right) \notag\\
&=& \!\!\!\!\!\!\!\! {2L_1L_2\over \lambda D}.
\end{eqnarray}
This result is in agreement with the classical DoF analysis derived in (\ref{eq:DoF_Shannon}). An alternative interpretation is by approximating the spatial frequency bandwidth $B(x')\approx B(0) = 2({2\over \lambda})\sin(\alpha(0)) \approx {4\over \lambda} {L_1\over 2D} = {2L_1\over \lambda D}$ for all $x'$, which yields $\text{SBP}_{G1} \approx L_2 B(0) = {2L_1 L_2\over \lambda D}$. 
This interpretation of the SBP calculation in the Fresnel regime can be easily extended to geometry G3 (Fig. \ref{fig:G3}) by projecting the spatial frequency bandwidth $B(0)$ onto the line that crosses the origin and creates an angle $\theta$ with the $k_{x}$ coordinate, to obtain the following approximate formula, 
\begin{equation}
\text{SBP}_{G3} \approx {2L_1 L_2\over \lambda D} \cos(\theta).
\end{equation}
It is worth mentioning that the seminal work of Slepian on PSWFs and the study of the Fourier kernel has been applied in various engineering problems, in particular in the context of diffraction limited optics \cite{Optics}, and line-of-sight MIMO communications \cite{MIMO}.

\subsection{Multistatic Array in the Fresnel regime}
\label{sec:multi_Fresnel}
In this subsection, we investigate multistatic imaging arrays under the the Fresnel approximation. The Fresnel diffraction integral in (\ref{eq:fresnel}) for an arbitrary Tx/Rx pair can be further simplified as,
\begin{eqnarray}
s(x_{tx}, x_{rx}) \!\!\!\!\!\! &&\approx e^{-j2kD} e^{-j{k\over 4D}(x_{tx} - x_{rx})^2} \times \notag\\
&&  \int\limits_{x'} \bgamma(x') e^{-j{k\over D}(x' - x_{mid})^2} dx',
\label{eq:multi_Fresnel}
\end{eqnarray}
where $x_{mid} \triangleq {1\over 2}(x_{tx} + x_{rx})$ represents the {\it midpoint} of the Tx/Rx pair. Comparing (\ref{eq:multi_Fresnel}) with (\ref{eq:mono_Fresnel}), we can see that the information being captured by the multistatic Tx/Rx pair is equivalent to that of a monostatic transceiver located at $x_{mid}$. Following similar lines of reasoning as we did for the analysis of monostatic arrays in Subsection \ref{sec:mono_Fresnel}, we can identify the number of the degrees of freedom for the integral kernel in (\ref{eq:multi_Fresnel}) by,
\begin{equation}
\text{Fresnel}_{dof} = \Delta x' \Delta f = L_2 ({k\Delta x_{mid}\over \pi D}) = {2L_1L_2\over \lambda D}.
\label{eq:DoF_Shannon_multi}
\end{equation}

Fresnel regime analysis of 1D multistatic architecture through PSWFs theory has also appeared in \cite{Pierri_multistatic}.  In the derivation of (\ref{eq:DoF_Shannon_multi}), we have used the observation that $x_{mid}$ is restricted to the aperture of the imaging system, hence the extent of its admissible values is given by $\Delta x_{mid} = L_1$. This result agrees with our previous observation through the space-bandwidth product analysis in Section \ref{sec:SBP}, where we established that the SBPs achieved by monostatic and multistatic architectures are equal to each other. 

The approximate integral equation in (\ref{eq:multi_Fresnel}) has  significant practical implications for multistatic array design. Most importantly, (\ref{eq:multi_Fresnel}) implies that a multistatic architecture can be replaced with an {\it effective monostatic} array, by placing a monostatic transceiver element at the midpoints of every Tx/Rx pair. Let $a_{tx}(x)$ and $a_{rx}(x)$ denote aperture functions that encode the locations of the transmitter and receiver elements, respectively, as follows:
\begin{eqnarray}
a_{tx}(x) = \sum_{i = 1}^{N_{tx}} \delta(x - x_{tx}(i)), \notag\\
a_{rx}(x) = \sum_{j = 1}^{N_{rx}} \delta(x - x_{rx}(j)), 
\label{eq:aperture_functions}
\end{eqnarray}
where $\delta(x)$ is the Dirac delta function.
By definition, the effective monostatic array is given by,
\begin{equation}
\label{eq:effective_sums}
a_{eff}(x) \triangleq \sum_{i = 1}^{N_{tx}} \sum_{j = 1}^{N_{rx}} \delta\left(x - \left({ x_{tx}(i) + x_{rx} (j)\over 2}\right)\right).
\end{equation}
As shown in Appendix \ref{app:effective_aperture}, $a_{eff}(x)$ can also be expressed as,
\begin{equation}
\label{eq:effective_conv}
a_{eff}(x) = a_{tx}(2x) \ast a_{rx}(2x),
\end{equation}
That is, the effective monostatic array is derived by {\it shrinking} the Tx and Rx aperture functions by a factor of $2$, followed by a {\it convolution} in the spatial domain. In the literature, the convolution expression for describing the effective monostatic array is mainly justified through {\it array factor} arguments (see Appendix \ref{app:effective_aperture} for a quick review), but the shrinkage step does not fall out of this approach \cite{Lockwood_effective_aperture, Ahmed_effective_aperture}. 

\begin{figure}[htbp]
\centering
\includegraphics[scale=0.43] {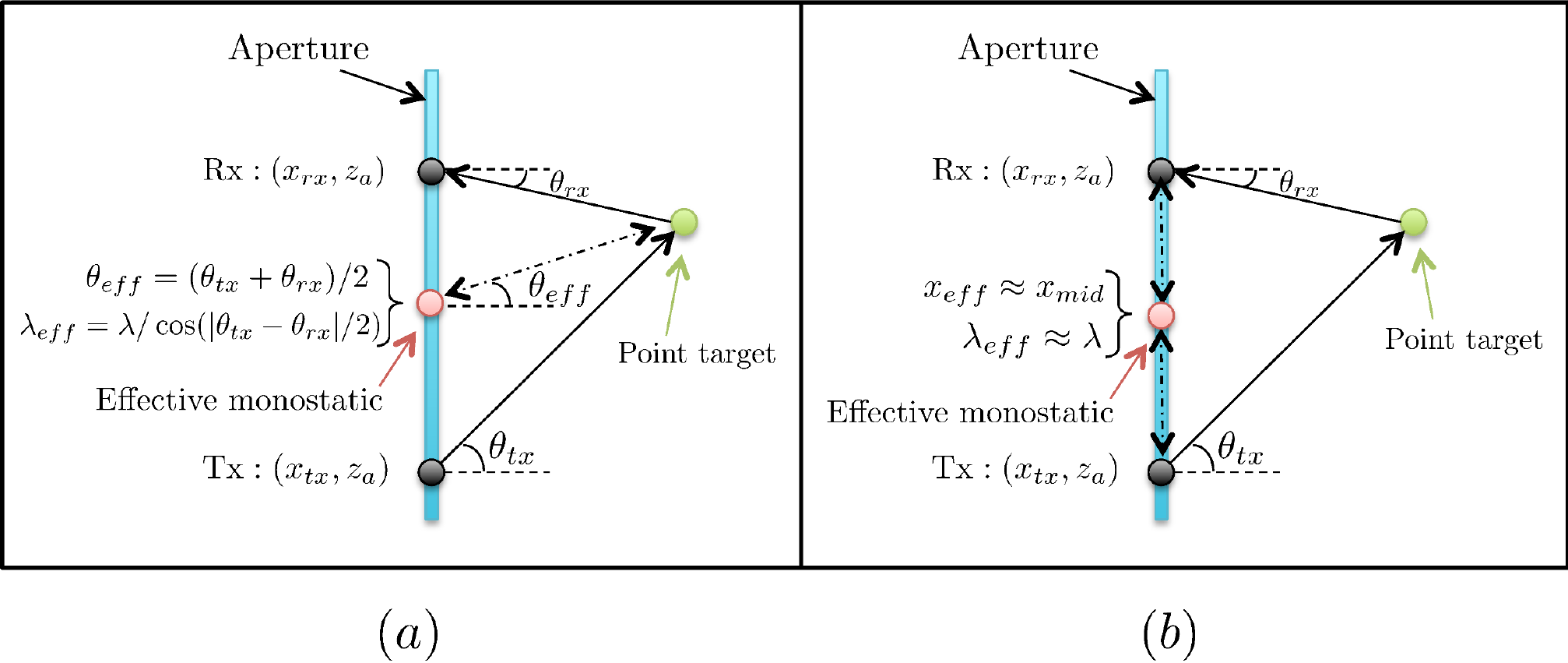}
  \caption{Effective monostatic element corresponding to a bistatic pair (a) using $k$-space analysis (valid under the Born approximation), and (b) after Fresnel far field approximation.}
  \label{fig:multi_Fresnel} 
\end{figure}

In Section \ref{sec:k_space_spectrum}, we showed that for a given point scatterer in the scene we can replace any spatially-separated Tx/Rx pair by a monostatic element that captures the exact same information. However, the constructed effective monostatic array is not generalizable to the entire scene, and depends on the location of the particular point scatterer being considered.  The Fresnel approximation leads to an effective monostatic that is independent from the scene, and provides a systematic approach for designing and analyzing multistatic architectures in the Fresnel regime. Figure \ref{fig:multi_Fresnel} summarizes the construction of the effective monostatic array using the the Fresnel approximation, compared to the solution provided by the $k$-space analysis. 

\begin{remark}
Our SBP analysis reveals that a multistatic array with an infinite number of array elements does not lead to an improvement in the number of DoF compared to that of a monostatic array for 1D imaging scenario. However, the multistatic approach is attractive when we wish to design an array with a small number of transceivers to capture as many degrees of freedom as possible. The effective array argument suggests that it is possible to realize a {\it dense} effective array by intelligent co-design of {\it sparse} transmitter and receiver arrays. Moreover, in the presence of noise, deploying a multistatic architecture leads to an improvement in the signal to noise ratio (SNR) compared to a monostatic array with the same number of elements.
\end{remark}

\section{Image Formation and Resolution Analysis}
\label{sec:reconstruction}

Image formation techniques aim to reconstruct the reflectivity function of the scene from the measured data, by solving the inverse scattering problem \cite{Chew_book}. Inverse scattering has a wide variety of applications in fields such as medical imaging, non-destructive testing, optics, and remote sensing \cite{Chew_book, Louis_medical, optical_tomography}. In general, inverse scattering problems are ill-posed due to the non-trivial nullspace of the imaging system, which implies that the reflectivity function satisfying our measurement equations is not unique.  In this section, we study the implications of DoF analysis for image reconstruction, and analyze the achievable image resolution.

\subsection{Pseudoinverse Reconstruction}
Since the number of DoF of the imaging systems is finite, we can rewrite equation (\ref{eq:Xi_decomposition}) as,
\begin{equation}
\label{eq:Xi_decomposition_finite}
\bs \approx \sum_{i=1}^{\text{DoF}} \sigma_{i} \phi_{i} \langle \bgamma , \psi_{i}\rangle_{\Psi}.
\end{equation}
This regularization procedure is known as Truncated SVD (TSVD) \cite{Pierri_Resolution_2D, Bertero_book}, where we only keep the nonzero singular values with a significant contribution to the measured data, and set the remaining singular values to zero. Assuming noiseless measurements, the ``best approximate'' solution to the image formation problem is given by the minimum $\ell^2$-norm estimate of the scene reflectivity function consistent with the data. This solution is obtained by computing $\Xi^{\dagger}\bs$, where $\Xi^{\dagger}: \Phi \to \Psi$ is the Moore-Penrose pseudoinverse (PINV) of $\Xi$. The reconstructed image is given by,
\begin{equation}
\label{eq:svd_reconstruction}
\hat{\bgamma}_{pinv} = \sum_{ i = 1}^{\text{DoF}} {1\over \sigma_i} \psi_{i} \langle \bs , \phi_i \rangle_{\Phi}.
\end{equation}
Combining (\ref{eq:Xi_decomposition_finite}) and (\ref{eq:svd_reconstruction}), gives us,
\begin{equation}
\label{eq:gamma_hat_tsvd}
\hat{\bgamma}_{pinv} = \sum_{ i = 1}^{\text{DoF}} \psi_{i}  \langle \bgamma , \psi_{i}\rangle_{\Psi},
\end{equation}
that is, the image is formed by projecting the original reflectivity function $\bgamma$ onto the subspace spanned by $\{\psi_{i}: i = 1, \dots, \text{DoF}\}$. Note that the reconstruction error ($\bgamma - \hat{\bgamma}_{pinv}$) lies approximately in the null-space of $\Xi$, i.e., $\Xi(\bgamma - \hat{\bgamma}_{pinv}) \approx 0$, and is therefore not observed through the imaging system. See \cite{Cakoni_book} for more details on TSVD-based pseudoinverse operation.


\subsection{Matched-Filter/Back-Propagation Reconstruction}
The standard and classical method for image reconstruction is based on applying the Hermitian adjoint operator to the measured data:
\begin{equation}
\label{eq:adj_reconstruction}
\hat{\bgamma}_{adj} = \Xi^{\ddagger}\bs,
\end{equation}
where $\Xi^{\ddagger}: \Phi \to \Psi$ denoted the adjoint of $\Xi$. This procedure is also known as {\it Matched-Filtering} (MF) \cite{Nolan_SAR}, and {\it Back-Propagation} algorithm \cite{Ahmed_backpropagation}. The integral operation corresponding to (\ref{eq:adj_reconstruction}) is identified by, 
\begin{equation}
\label{eq:adj_integral}
\hat{\bgamma}_{adj}(x'' , z'') = \int\limits_{\mathcal{B}} \xi^{*}(x_{tx},x_{rx}, x'', z'') \bs(x_{tx}, x_{rx}) dx_{tx} dx_{rx},
\end{equation}
where $\xi^{*}$ is the complex conjugate of $\xi$ (It is easy to verify that $\langle \Xi \bgamma , \bs \rangle_{\Phi} = \langle \bgamma , \Xi^{\ddagger} \bs\rangle_{\Psi}$). Combining (\ref{eq:born}) and (\ref{eq:adj_integral}) gives us
\begin{equation}
\label{eq:integral_using_kappa}
\hat{\bgamma}_{adj}(x'' , z'') = \int\limits_{\mathcal{A}} \kappa(x'',z'',x',z') \bgamma(x',z') dx'dz',
\end{equation}
where $\kappa$ corresponds to the compact self-adjoint linear operator defined by,
\begin{eqnarray}
\kappa(x'',z'',x',z') &=& \notag\\ 
\int\limits_{\mathcal{B}} && \!\!\!\!\!\!\!\!\!\!\!\!\!\!\!\!  \xi^{*}(x_{tx},x_{rx}, x'', z'') \xi(x_{tx},x_{rx}, x', z') dx_{tx} dx_{rx}. \notag\\
\end{eqnarray}
Using (\ref{eq:zeta_decomposition}), we can rewrite (\ref{eq:integral_using_kappa}) as
\begin{equation}
\label{eq:gamma_hat_adj}
\hat{\bgamma}_{adj} = \sum_{ i = 1}^{\infty} \sigma_{i}^{2} \psi_{i}  \langle \bgamma , \psi_{i}\rangle_{\Psi} \approx  \sum_{ i = 1}^{\text{DoF}} \sigma_{i}^{2} \psi_{i}  \langle \bgamma , \psi_{i}\rangle_{\Psi},
\end{equation}
This corresponds to projecting $\bgamma$ onto the subspace spanned by $\{\psi_{i}: i = 1, \dots, \text{DoF}\}$, while {\it weighting} the components of the projection by the square of the corresponding singular values. Note that if the singular values of the imaging system are approximately equal, $\sigma_i \approx \sigma, ~\forall i$, then $\hat{\bgamma}_{adj} \approx \sigma^2 \hat{\bgamma}_{pinv}$; that is, the image formed by the adjoint operator is just a scaled version of the output of Moore-Penrose pseudoinverse operator. 


\subsection{Resolution Analysis}
One of the most important performance metrics for any imaging systems is its resolution capability. Here we use the classical Rayleigh criterion and  associated reciprocal bandwidth arguments \cite{resolution_survey} to compare the performance of different reconstructions schemes for monostatic and multistatic arrays.  It follows from the uncertainty principle that a function's width in the spatial domain is inversely proportional to its width in the spatial frequency domain \cite{Zeev_book}. Hence, we use $1/B(x',z')$  as a benchmark measure of achievable resolution for a point scatterer located at $(x',z')$. We fix $L_1 = 15$cm, $L_2 = 10$cm, and $D = 40$cm for the numerical results of this subsection.

\begin{figure}[htbp]
\centering 
\subfigure[]{
\includegraphics[scale=0.19]{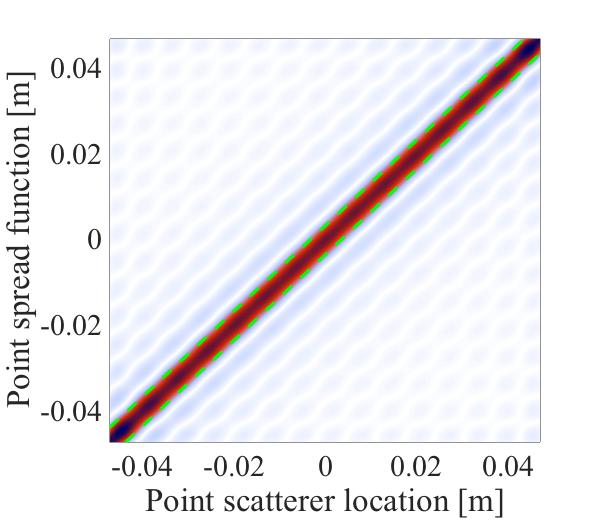}
}
\subfigure[]{
\includegraphics[scale=0.19]{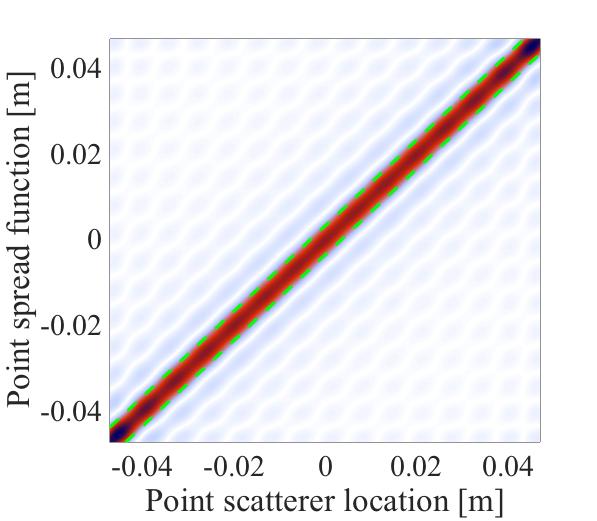}
}
\caption{PSFs with PINV reconstruction corresponding to geometry G1. We plot $|\hat{\gamma}_{pinv}(x'', z'' = 0)|$ with $x''$ along the vertical axis, and true point scatterer location $x'_{p}$ along the horizontal axis, for (a) monostatic, and (b) multistatic arrays. The reciprocal spatial frequency bandwidth $1/B(x',z')$, is depicted by the dashed lines. Figure is best viewed in electronic version.}
\label{fig:PSF_TSVD} 
\end{figure}

\begin{figure}[htbp]
\centering 
\subfigure[]{
\includegraphics[scale=0.19]{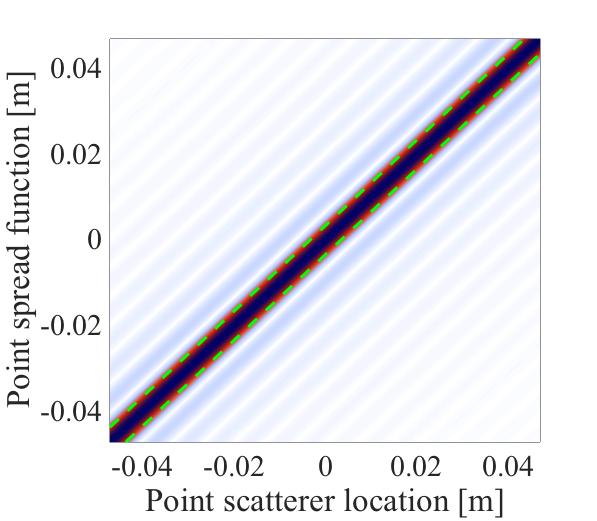}
}
\subfigure[]{
\includegraphics[scale=0.19]{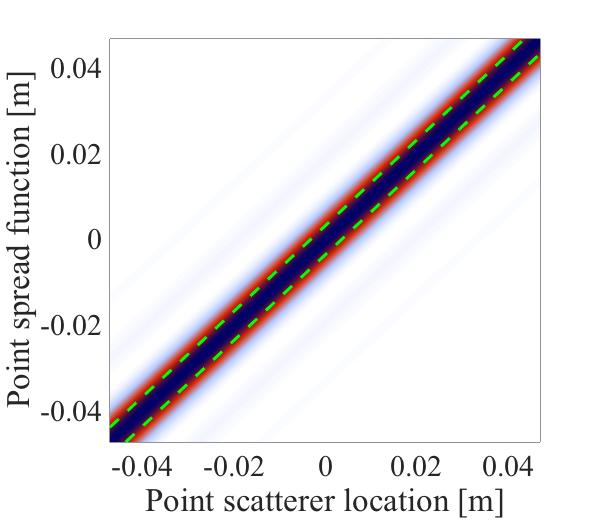}
}
\caption{PSFs with MF reconstruction corresponding to geometry G1. We plot $|\hat{\gamma}_{adj}(x'', z'' = 0)|$ with $x''$ along the vertical axis, and true point scatterer location $x'_{p}$ along the horizontal axis, for (a) monostatic, and (b) multistatic arrays. The reciprocal spatial frequency bandwidth $1/B(x',z')$, is depicted by the dashed lines. Figure is best viewed in electronic version.}
\label{fig:PSF_MF} 
\end{figure}

\begin{figure}[htbp]
\centering
\includegraphics[scale=0.25] {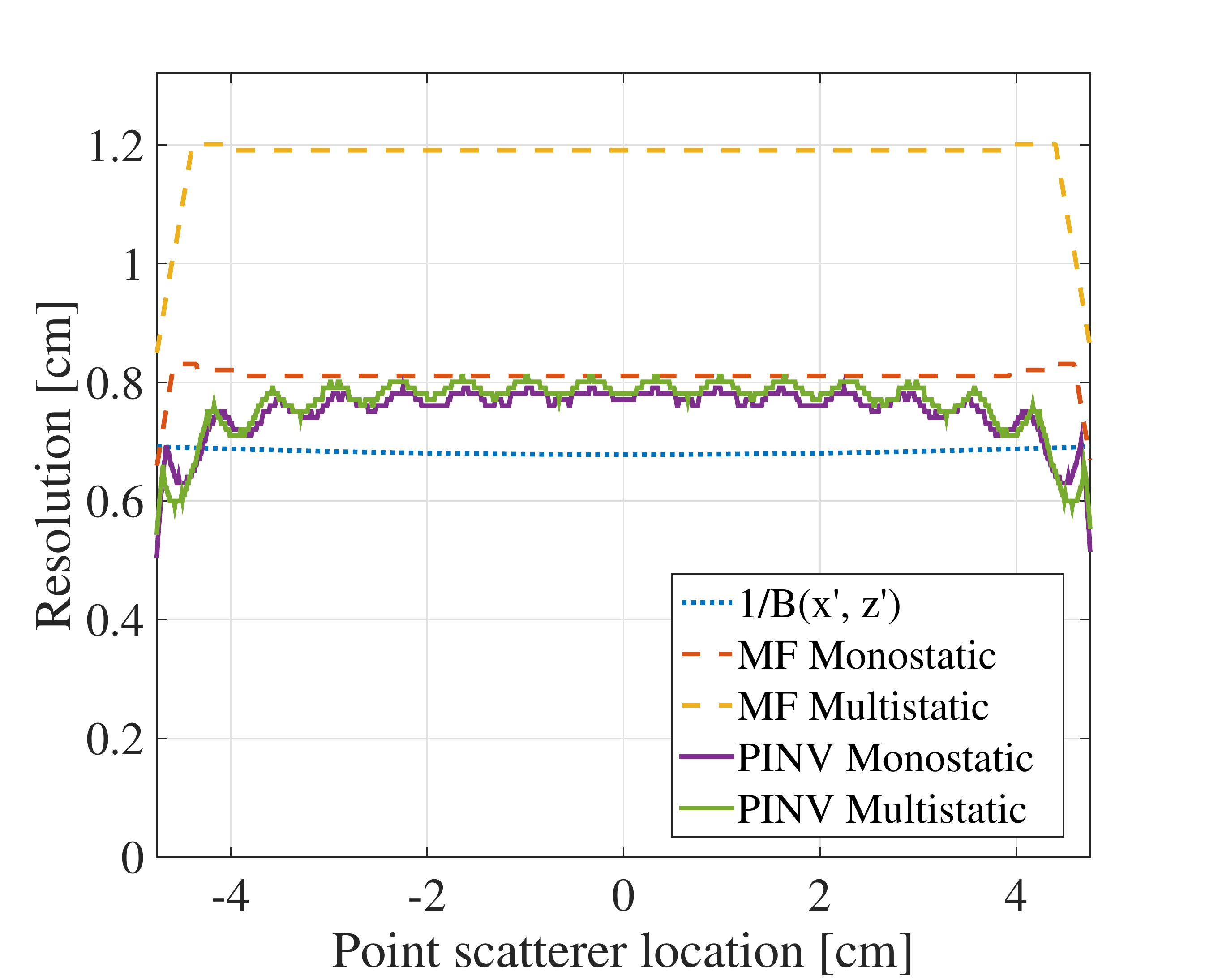}
  \caption{Resolution evaluated by 3dB beamwidth of the mainlobe of PSFs for geometry G1.}
  \label{fig:G1_resolution} 
\end{figure}

Based on the Rayleigh criterion, the resolution of an imaging system is defined by the spatial width of the reconstructed image corresponding to a point target, also known as point spread function (PSF). For a point target located at $(x'_{p}, z'_{p})$, we substitute $\bgamma(x', z') = \delta(x' - x'_{p})\delta(z' - z'_{p})$ in (\ref{eq:gamma_hat_tsvd}) and (\ref{eq:gamma_hat_adj}), to obtain PSFs corresponding to the pseudoinverse (Figure \ref{fig:PSF_TSVD}) and matched filter (Figure \ref{fig:PSF_MF}) reconstruction schemes, respectively. From (\ref{eq:integral_using_kappa}), it is evident that $\kappa(x'', z'', x'_{p}, z'_{p})$ also identifies the PSF corresponding to the matched filter scheme.  In order to quantify the achievable resolution for different scenarios, we evaluate the 3dB beamwidth of the mainlobe of the PSFs, and compare the results with $1/B(x',z')$. As depicted in Figure \ref{fig:G1_resolution}, for geometry G1, PINV reconstruction outperforms MF, and leads to a better resolution throughout the scene. More importantly, we see a significant resolution loss for the MF method for multistatic architecture. This can be partially explained by our previous observation that, if the variation across the significant singular values of the imaging system is large (which is shown to be the case for multistatic arrays based on our SVD computations), then MF reconstruction deviates from the optimal PINV operation.  

\begin{figure}[htbp]
\centering 
\subfigure[]{
\includegraphics[scale=0.25]{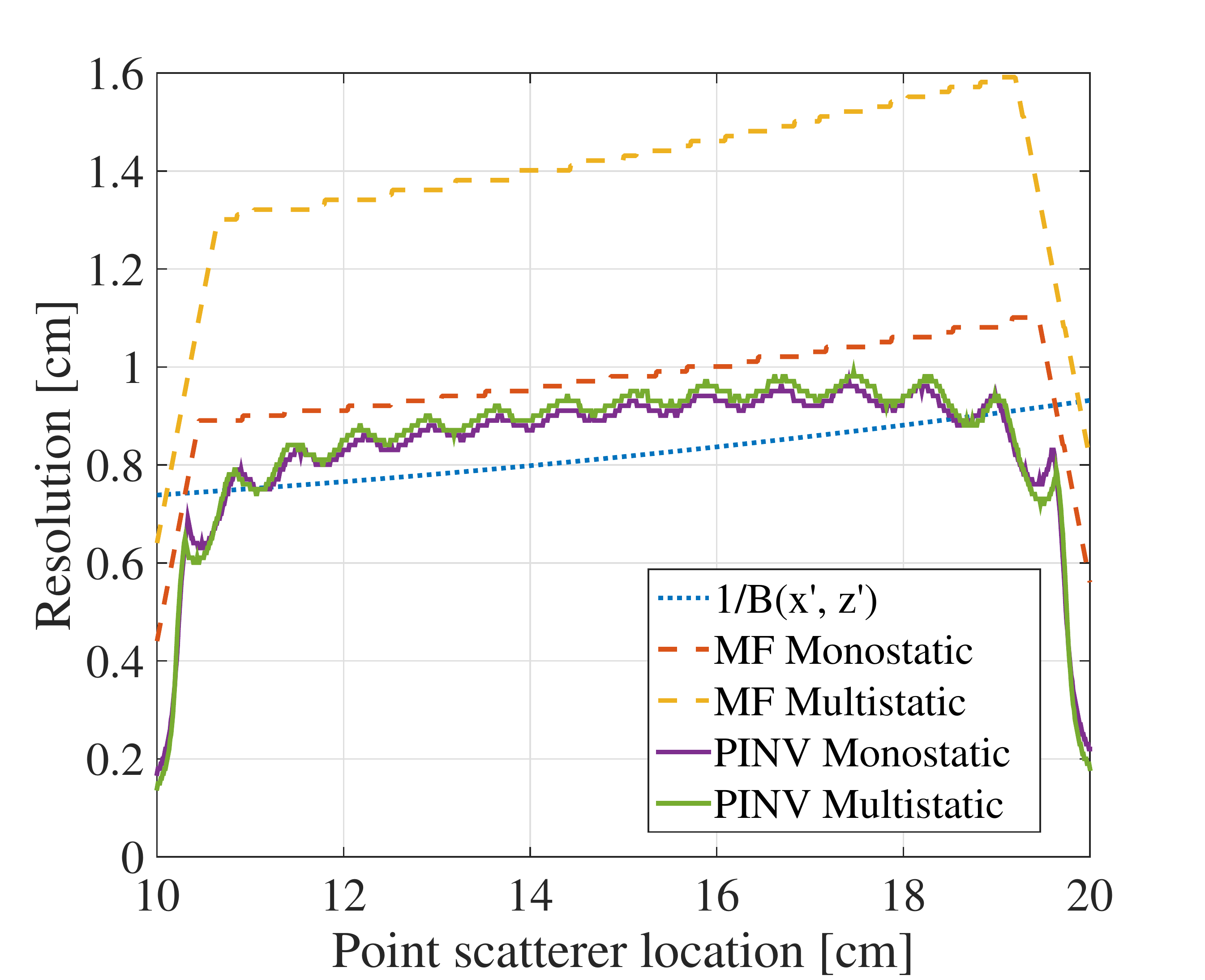}
}
\subfigure[]{
\includegraphics[scale=0.25]{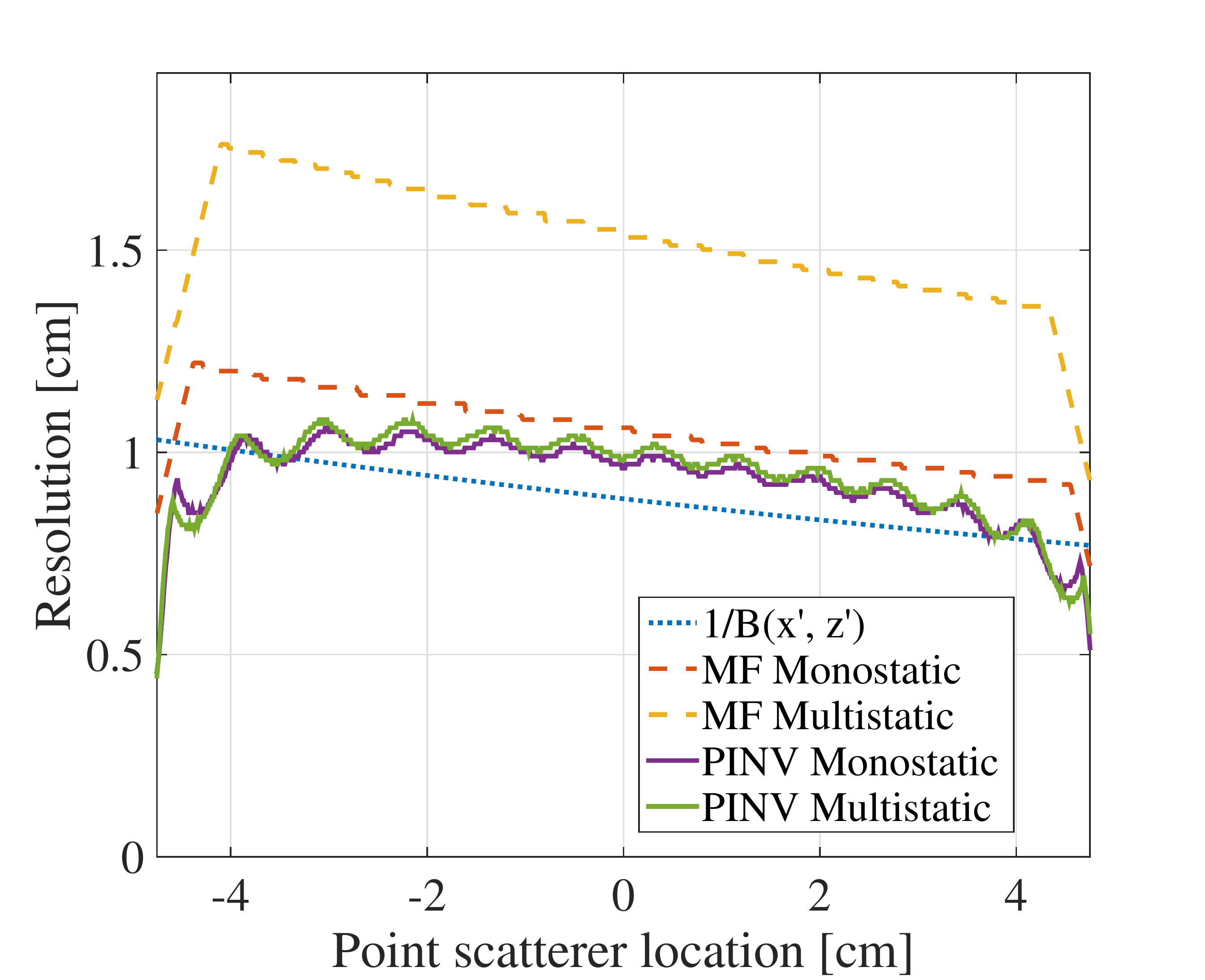}
}
\caption{Resolution evaluated by 3dB beamwidth of the mainlobe of PSFs for (a) geometry G2 with $t=15$cm, and (b) geometry G3 with $\theta = 40^{\circ}$.}
\label{fig:G2_G3_resolution} 
\end{figure}

We also evaluate the achievable resolution corresponding to geometries G2 and G3, with $t = 15$cm and $\theta = 40^{\circ}$, respectively. As depicted in Figure \ref{fig:G2_G3_resolution}-a for geometry G2, the resolution is inversely related to the distance of the point scatterer from the center of the aperture. Also, as shown in Figure \ref{fig:G2_G3_resolution}-b for geometry G3, rotation of the scene leads to an improvement in resolution for the point scatterers that are closer to the aperture, and a degradation for the ones that are farther. These results are intuitively appealing as well as theoretically justifiable through reciprocal bandwidth arguments.

\begin{remark}
The number of DoF and the SVD (including the singular values and the singular functions) corresponding to an imaging scenario depend on the scene extent $L_2$, whereas the resolution of a point scatterer and its spatial frequency bandwidth $B(x',z')$ are  independent of $L_2$. This leads us to an interesting observation: even though PINV reconstruction uses SVD for inversion (which is a global decomposition of the imaging scenario), its corresponding resolution limit closely follows the reciprocal bandwidth $1/B(x',z')$, which is a locally defined measure of accuracy in determining the location of the point scatterers. Thus, while the eigenmodes from the SVD are quite complex, and involve both the scene and the image, the capability of the imaging system is captured accurately via the stationary phase approximation underlying the $k$-space representation for each point in the scene.
\end{remark}


\section{Conclusions}
\label{sec:conclusions}
We have introduced the space-bandwidth product as a measure of predicting the number of DoF of 1D radar imaging systems under the Born approximation. Our approach extends to 2D planar geometries, but detailed analysis for such settings is deferred to future work. Our results are more general than prior analysis in the Fresnel regime, and thus provide insights for short-range scenarios where the Fresnel approximation breaks down. We have validated the accuracy of SBP in predicting the DoF by numerical evaluations of the singular decomposition of the imaging system in various scenarios. Our analysis reveals that, in terms of achievable degrees of freedom, there is no fundamental benefit in a multistatic architecture relative to a monostatic architecture. However, from a practical point of view, a key advantage of a multistatic architecture is that a large ``effective monostatic'' array can be synthesized using sparse transmit and receive arrays.
Our analytical framework also opens up the design space for new signal processing techniques that are capable of incorporating prior information into the image reconstruction procedure, and provides crucial insights on the achievable resolution. Further investigation of these techniques is an important topic for future work.
\section{Acknowledgement} 
This work was supported in part by Systems on Nanoscale Information fabriCs (SONIC), one of the six SRC STARnet Centers, sponsored by MARCO and DARPA, and in 
part by the National Science Foundation by grant CNS-1518812.

\begin{appendices}

\section{Proof of sum rule}
\label{app:sum_rule}
The Hilbert-Schmidt norm of the operator $\Xi$ is given by,

\begin{eqnarray}
&||\Xi||_{HS}^2& = \iint\limits_{\mathcal{A}\mathcal{B}} |\xi(x_{tx}, x_{rx}, x',z')|^{2} dx_{tx}dx_{rx}dx' dz' \notag\\
&\stackrel{(a)}{=}& \iint\limits_{\mathcal{A}\mathcal{B}} \sum_{i = 1}^{\infty} \sigma_{i} \phi_{i}(x_{tx},x_{rx}) \psi^{*}_{i}(x',z') \times \notag\\ 
&&\sum_{j = 1}^{\infty} \sigma_{j} \phi^{*}_{j}(x_{tx},x_{rx}) \psi_{j}(x',z') dx_{tx}dx_{rx}dx' dz' \notag\\
&\stackrel{(b)}{=}&  \sum_{i = 1}^{\infty}  \sum_{j = 1}^{\infty} \sigma_i \sigma_j \int\limits_{\mathcal{B}} \phi_{i}(x_{tx},x_{rx}) \phi^{*}_{j}(x_{tx},x_{rx}) dx_{tx}dx_{rx}\notag\\
&& \int\limits_{\mathcal{A}}  \psi^{*}_{i}(x',z') \psi_{j}(x',z') dx' dz' \notag\\
&\stackrel{(c)}{=}&  \sum_{i = 1}^{\infty}  \sum_{j = 1}^{\infty} \sigma_i \sigma_j  \delta(i - j) =  \sum_{i = 1}^{\infty} \sigma_i^2,
\end{eqnarray}
where $(a)$ is by the SVD in (\ref{eq:zeta_decomposition}), $(b)$ is by changing the order of integrations and summations, and $(c)$ follows from the fact that the sets of singular functions $\{\psi_{i}\}_{i\in\mathbb{N}}$ and $\{\phi_{i}\}_{i\in\mathbb{N}}$ are orthonormal bases for $\Psi$ and $\Phi$, respectively. This completes the proof for the sum rule equality in (\ref{eq:sum_rule}).

\section{Proof of Theorem 1}
\label{app:proof}
 The following lemmas will be used in the proof.

\begin{lemma}
\label{lemma_1}
Let $\mathcal{T}_1$ and $\mathcal{T}_2$ be two sets of points in 2D space, with $\mathcal{T}_1 \subseteq \mathcal{T}_2$. Then, $\mathcal{I}_{l}(\mathcal{T}_1) \subseteq \mathcal{I}_{l}(\mathcal{T}_2)$.

\noindent{\bf Proof:} For any $p_1 \in \mathcal{I}_{l}(\mathcal{T}_1)$, $\exists~ p_2 \in \mathcal{T}_1$, such that $p_1 = \mathcal{I}_{l}(p_2)$. Since $\mathcal{T}_1 \subseteq \mathcal{T}_2$, we have $p_2 \in \mathcal{T}_2$. Therefore, $p_1 \in \mathcal{I}_{l}(\mathcal{T}_2)$. This completes the proof. \hfill $\square$ 
\end{lemma}

\begin{figure}[htbp]
\centering
\includegraphics[scale=0.5] {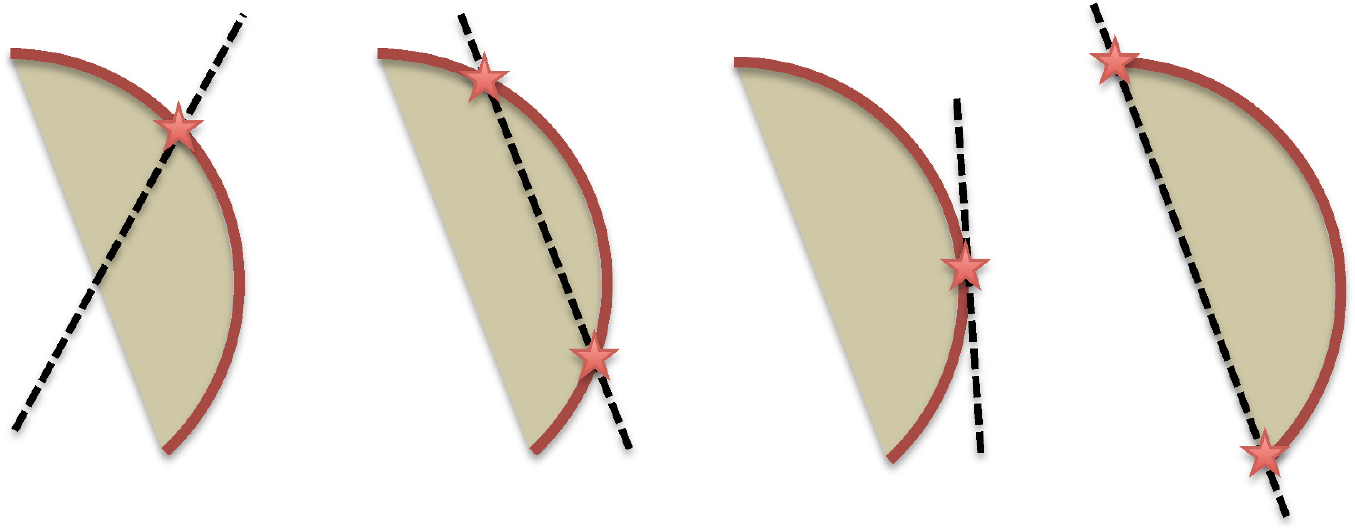}
  \caption{Proof of Lemma \ref{lemma_2}. The intersection of the dashed line with the circular segment has at least one point on the arc, depicted by the star symbols.}
  \label{fig:lemma_2} 
\end{figure}

\begin{lemma}
\label{lemma_2}
The intersection of a circular segment with a line in 2D space, is either the empty set, or it contains at least one point from the arc boundary of the circular segment. The proof is simple and shown by Figure \ref{fig:lemma_2}.  
\end{lemma}

{\bf Proof of Theorem 1:} From (\ref{eq:T_subset_T}) we have $\Tmono \subseteq \Tmulti \subseteq \text{conv}(\Tmono)$. By Lemma \ref{lemma_1}, 
\begin{equation}
\label{eq:I_subset_I}
\mathcal{I}_{l}(\Tmono) \subseteq \mathcal{I}_{l}(\Tmulti) \subseteq \mathcal{I}_{l}(\text{conv}(\Tmono)).
\end{equation} 
Note that $\Tmono$ is an arc of a circle of radius $2k$, and its convex hull forms the corresponding circular segment. For any point $p_1 \in \mathcal{I}_{l}(\text{conv}(\Tmono))$, define the inverse image  as $\mathcal{I}_{l}^{-1}(p_1) \triangleq \{p_2 \in  \text{conv}(\Tmono): \mathcal{I}_{l}(p_2) = p_1\}$. It is easy to see that $\mathcal{I}_{l}^{-1}(p_1)$ is the intersection of $\text{conv}(\Tmono)$ with a line passing through $p_1$ and perpendicular to $l$. By Lemma \ref{lemma_2}, $\mathcal{I}_{l}^{-1}(p_1)$ includes at least one point $p_2\in \Tmono$ (i.e., in the arc boundary of $\text{conv}(\Tmono)$). Therefore, $p_1 = \mathcal{I}_{l}(p_2) \in \mathcal{I}_{l} (\Tmono)$, so that $\mathcal{I}_{l}(\text{conv}(\Tmono)) \subseteq  \mathcal{I}_{l} (\Tmono)$. Combining this with ($\ref{eq:I_subset_I}$), we obtain
\begin{equation}
\mathcal{I}_{l}(\Tmono) = \mathcal{I}_{l}(\Tmulti) = \mathcal{I}_{l}(\text{conv}(\Tmono)).
\end{equation}
This completes the proof. \hfill $\square$ 

\section{Effective Aperture Concept}
\label{app:effective_aperture}
Effective aperture (also known as virtual array) is a widely used technique for the design and analysis of multistatic arrays in far field regimes \cite{Lockwood_effective_aperture, Ahmed_effective_aperture}. Based on this approach, one can derive an equivalent monostatic array for any given multistatic architecture by convolving the transmit and receive aperture functions \cite{Pal_nested_array} (defined in (\ref{eq:aperture_functions})). The classical analysis of the effective aperture relies on the notion of {\it array factor} (radiation pattern of the array in far field) defined by the Fourier transform of the aperture functions,
\begin{eqnarray}
P_{tx}(g) = \int_{\mathbb{R}} a_{tx}(x) e^{-j2\pi gx} dx, \notag\\
P_{rx}(g) = \int_{\mathbb{R}} a_{rx}(x) e^{-j2\pi gx} dx,
\end{eqnarray}
where $g\triangleq \sin(\theta)$, $\theta$ being the angle measured from the perpendicular to the array. The two-way array factor is given by the product of the transmit radiation pattern $P_{tx}(\cdot)$, and receive radiation pattern $P_{rx}(\cdot)$, 
\begin{equation}
P_{eff}(g) = P_{tx}(g).P_{rx}(g).
\end{equation}
Equivalently, the effective aperture is given by the convolution of the Tx and Rx aperture functions, i.e., $a_{tx}(x) \ast a_{rx}(x)$. While this analysis is intuitively pleasing, it does not capture the {\it shrinkage} operation described in Subsection \ref{sec:multi_Fresnel}. Here, we show that the shrinkage and convolution operations in (\ref{eq:effective_conv}) are indeed consistent with the effective monostatic array (\ref{eq:effective_sums}) derived from the Fresnel approximation;
\begin{eqnarray}
&&a_{tx}(2x)\ast a_{rx}(2x) \notag\\
&&=\sum_{i=1}^{N_{tx}} \delta\left(x - {x_{tx}(i) \over 2}\right) \ast \sum_{j=1}^{N_{rx}} \delta\left(x - {x_{rx}(j)\over 2}\right) \notag\\
&&=\int_{p}~ \sum_{i=1}^{N_{tx}} \delta\left(x - {x_{tx}(i) \over 2} - p\right) . \sum_{j=1}^{N_{rx}} \delta\left(p - {x_{rx}(j)\over 2}\right) dp \notag\\
&& \stackrel{(a)}{=} \sum_{i = 1}^{N_{tx}} \sum_{j = 1}^{N_{rx}} \delta\left(x - \left({ x_{tx}(i) + x_{rx} (j)\over 2}\right)\right) = a_{eff}(x),
\end{eqnarray}
where $(a)$ follows from the sifting property of the delta function.

\end{appendices}

\bibliographystyle{ieeetr}
\bibliography{Ref}

\end{document}